\input{epsf}

\documentstyle[preprint,aps,eqsecnum]{revtex}
\tightenlines
\begin{document}
\setcounter{page}{0}
\preprint{UW/PT-99-09}
\title
{ANOMALOUS PSEUDOSCALAR-PHOTON VERTEX IN AND OUT OF EQUILIBRIUM}
\author
{\bf S. Prem Kumar$^{(a)}$, Daniel Boyanovsky$^{(b,c)}$, Hector J. de
Vega$^{(c,b)}$\\and Richard Holman$^{(d)}$ }
\address
{\it (a) Department of Physics, University of Washington, Seattle,
WA 98195, U.S.A.\\
(b) Department of Physics and Astronomy, University of 
Pittsburgh, Pittsburgh  PA. 15260, U.S.A \\ (c)
LPTHE\footnote{Laboratoire Associ\'{e} au CNRS UMR 7589.}  Universit\'e Pierre
et Marie Curie (Paris VI) et Denis Diderot (Paris VII), Tour 16, 1er. \'etage,
4, Place Jussieu 75252 Paris, Cedex 05, France  \\
(d) Department of Physics, Carnegie-Mellon University, Pittsburgh, 
PA. 15213, U.S.A.} 
\date{\today}
\maketitle 
\begin{abstract}
The anomalous pseudoscalar-photon vertex is studied in real time in and out of
equilibrium in a constituent quark model. The goal is to understand the
in-medium modifications of this vertex, exploring the possibility of
enhanced isospin breaking by electromagnetic effects as well as the
formation of 
neutral pion condensates in a rapid chiral phase transition in peripheral,
ultrarelativistic heavy-ion collisions. 
In {\em equilibrium} the effective vertex is
afflicted by infrared and pinch singularities that require hard thermal
loop (HTL) and width corrections of the quark propagator. 
The resummed effective equilibrium 
vertex vanishes near the chiral transition in the chiral limit. 
In a strongly {\em out of equilibrium} chiral phase transition we find that the chiral
condensate drastically modifies the quark propagators and 
the effective vertex. The ensuing
dynamics for the neutral pion results in a potential enhancement of
isospin breaking and the formation of $\pi^0$ condensates. 
While the anomaly equation and the axial Ward identity are not 
modified by the medium in or out of equilibrium, the effective {\em real-time}
pseudoscalar-photon vertex is sensitive to low energy physics. 
\end{abstract} 

\section{Introduction}

The forthcoming ultrarelativistic heavy ion colliders, RHIC at BNL and
LHC at CERN will allow us to probe the physics of the quark gluon plasma, 
the hadronization and the chiral phase transitions, providing for
the first time, experimental access to a phase transition in particle
physics as predicted by QCD \cite{QCD}. Current experiments at CERN SPS are
seeking manifestations of these phase transitions with the large acceptance
detectors NA49 and WA98, while the large acceptance detector capabilities at
RHIC will allow an event-by-event analysis of {\em both} hadronic and
electromagnetic signals.
Currently WA98 at SPS-CERN studies fluctuations in the neutral to charged pion
ratios on an event-by-event 
basis \cite{wa98}, and previously the MINIMAX collaboration at Fermilab
studied similar signals\cite{fermilab}. These experiments are
particularly important in the detection of disoriented chiral
condensates (DCC) which are coherent regions of 
chirally misaligned pion condensates, proposed and discussed by many 
authors \cite{dcc}, originally within the context of CENTAURO events in cosmic
ray experiments\cite{centauro} but more recently as a possible signal for 
the chiral phase transition if it occurs out of equilibrium
\cite{moredcc,nosdcc,raja,photop}. 
With the turning on of the above mentioned experiments, with their capacity to
analyze data from ultrarelativistic heavy-ion collisions on an event-by-event
basis, a theoretical
exploration of various signals of the chiral phase transition is in order. 

The axial anomaly \cite{abj} has long been a source of extremely interesting
physics, ranging from the explanation of the $\pi^0 \rightarrow \gamma \gamma$ decay to
understanding the mass splitting between the $\eta$ and the $\eta'$
pseudoscalar mesons \cite{t'hooft}.
In this article we focus on the equilibrium and nonequilibrium aspects of the
{\em anomalous} electromagnetic interaction of the {\em neutral pion} to
investigate the possibility of experimentally important signals arising from
this interaction and its modifications in a medium.
In particular, since we expect that the collision region will contain
a large amount of energy density, a finite temperature study as well as a study
of the effects of the axial anomaly in out-of-equilibrium situations will prove important
for our understanding of forthcoming experimental results.

Detailed studies in equilibrium have found that the anomalous Ward identity is independent of
temperature and chemical potential\cite{gomez} and that in chiral perturbation theory, valid for
$T<<T_c$ the amplitude for neutral pion decay into two photons is independent of temperature.

However, in a series of recent articles, Pisarski \cite{rob} and Pisarski, Trueman and Tytgat
\cite{robtyt} have provided a detailed analysis of the chiral anomaly and
the anomalous {\em vertex} at finite temperature using the Euclidean formulation and
concluded,  that while the amplitude
for the neutral pion decay into two photons at zero temperature is directly
proportional to the coefficient of the axial anomaly (given by the anomalous
Ward identity), at finite temperatures this relationship no longer holds. 
This result was anticipated in \cite{itomuller} where it was pointed out that at
finite temperature there is a new vector (the four-velocity of the heat bath)
which leads to a more complicated tensor structure for the triangle diagram that
yields the anomalous amplitude and invalidates the tensor analysis leading 
to the Sutherland-Veltman theorem \cite{suth}.

Pisarski \cite{rob} also pointed out that near the critical
temperature for the chiral phase transition, the amplitude for
$\pi^0\rightarrow\gamma\gamma$ {\em vanishes}. This conclusion was based on an
analysis of the corresponding 3-point vertex with {\em spacelike} external
photons, using the imaginary time formalism. Gelis \cite{gelis} studied the
anomalous decay of the neutral pion into two photons at finite temperature and
pointed out some subtleties of the triangle diagram in the limit of soft
external momenta, and
reconciled certain discrepancies between the results of \cite{rob,robtyt} and
those of \cite{gnayak}.
A study in the linear sigma model at finite temperature and chemical potential {\em in equilibrium}
was anticipated in\cite{gomez2} with similar results to those found in\cite{gnayak,gelis}.

In this work we apply the {\em real time} formulation of quantum field theory
to the linear sigma model (LSM) coupled to constituent quarks and photons and 
calculate the effective coupling between the $\pi^0$ and photons both at finite
temperature as well as out of equilibrium. This allows us to obtain far more
information from the system than could be found from calculations of suitably
defined decay amplitudes (for e.g. \cite{gelis});
in particular, we will be able to construct a real time equation of motion for
the neutral pion condensate, and make contact with recent proposals of
DCC formation \cite{dcc,moredcc,nosdcc,raja,photop} via anomalous 
electromagnetic interactions \cite{asamuller}.

We first focus our attention on the equilibrium case in the region near the
chiral phase transition where the pions and constituent quarks are becoming
massless. We find that in the  
limit of soft photon momenta near the phase transition two different
types of divergences appear: infrared divergences in the chirally
restored phase associated with the masslessness of 
the quarks in the triangle diagram, and
pinch singularities that arise solely from the finite temperature
contributions. The infrared singularities near the chiral limit are the source
of discrepancy between the work in \cite{rob,robtyt} performed in the imaginary
time formulation and that of \cite{gnayak} in the real time version. This
discrepancy and its resolution has been recently discussed in \cite{gelis}. 
The pinch singularities have a very different origin. For photons of very
small virtuality, the quarks that are present in the medium and that can
contribute to the triangle diagram, can propagate on-shell. If the quarks are
treated as bare objects, then these on-shell states in the thermal bath will
propagate without damping and a pinch singularity will appear as a consequence
of these undamped on-shell intermediate states. 

Both singularities require that the internal quark propagators be dressed by
self-energy corrections, giving them a chirally symmetric mass as well as a
width of thermal origin. In the Yukawa theory, the fermion receives Hard
Thermal Loop (HTL) corrections to its self-energy \cite{gelis,thoma} when
its momentum $ <<g\,T $ where $ g $ is the Yukawa coupling and $ T $ the
temperature. As long as the quark mass derived from the tree-level
Yukawa interaction, $ m_q>>g\,T $, HTL's are 
unimportant as far as the real part of the fermion self-energy is concerned. 
However in the critical region in the strict chiral limit, the constituent
quark mass $ m_q $ vanishes and the contribution of hard thermal loops
(HTL) becomes important. The pinch singularities require a further
self-energy resummation that incorporates the on-shell width of the
thermal excitations; this goes beyond the HTL contribution. 

We will argue that both the hard thermal loop correction to the self-energy as
well as the width are dominated by scalar and pseudoscalar exchange.
Furthermore, we find that the quark width in the medium arises from the {\em
decay} of the scalar into quark-antiquark pairs as found previously 
in \cite{shang} within a different context and from collisional
contributions in 
the HTL limit \cite{thoma}.  We compute these corrections to the quark
propagators explicitly in the equilibrium case, and we find, in agreement with
\cite{rob,robtyt,gelis}, that the pion-photon vertex vanishes in the strict
chiral limit near the critical temperature. $\pi^0\rightarrow\gamma\gamma$
at finite temperature is, however, of limited relevance for heavy-ion collisions since
the typical lifetime of the neutral pion is several orders of magnitude
larger than that of the fireball (the latter being $\approx 50$fm/c).
The $\eta$ meson, however, has a much shorter lifetime $\approx 1.2$ KeV and 
finite temperature effects may be of importance for the process 
$\eta\rightarrow\gamma\gamma$ which occurs with a branching ratio of
40$\%$. Finite temperature effects 
could also be of importance in the cooling of hot and dense stars 
\cite{gnayak}, and in certain models of baryogenesis \cite{dine}. 

Our interest in the {\em nonequilibrium} calculation of the pion-photon anomalous
vertex comes in part from a recent proposal in \cite{asamuller}. It is argued
there that in the presence of strong electromagnetic fields, such as might be
found in peripheral collisions between heavy nuclei \cite{baur}, the
anomalous coupling 
might enhance DCC formation along the the neutral pion direction.
To quantify the effect of strong EM fields on DCC formation, note that the
coupling of the neutral pion to the electromagnetic field is given by 
$\sim g_{\pi \gamma \gamma} \pi_0 F_{\mu \nu}\tilde{F}^{\mu \nu}$.  In
peripheral 
collisions, the strong electromagnetic fields give rise to a {\em
semiclassical} contribution to $F_{\mu \nu}\tilde{F}^{\mu \nu} \propto
\vec{E}\cdot\vec{B} \propto Z^2e^2 b/R^6$ with $b$ the impact
parameter \cite{asamuller}. Therefore, for large impact parameters (very
peripheral collisions $b >> 2R$) and heavy nuclei the anomalous
interaction of the 
neutral pion with the electromagnetic field could induce large neutral pion
condensates. A further interesting scenario was investigated in
\cite{photop}, namely the possibility of the anomalous vertex inducing photon
production via parametric amplification during the chiral phase transition.
These investigations studied  the phenomenological consequences of the
anomaly by using the {\em vacuum} form for the vertex
$g_{\pi\gamma\gamma}\sim\alpha/f_\pi$. 

But our previous comments clearly suggest that the vertex {\em is not}
protected from corrections by the medium and the strength of the
interaction should be drastically modified by the 
nonequilibrium state, as is the case at finite temperature.
Our aim therefore is : i) to study the nature of these modifications and provide a
detailed construction of the vertex in {\em real time}, and 
ii) to provide a quantitative
assessment of the potential phenomenological consequences of 
these in-medium corrections. In this paper we deal with the first part of the
program, i.e. the more formal aspects of setting up a framework for studying
the out-of-equilibrium interactions and 
relegate to a forthcoming article, the
phenomenological and numerical implications of these modifications.

In the nonequilibrium case we assume that the initial state is some thermal
state where the chiral order parameter has a small 
amplitude set by the explicit chiral 
symmetry breaking term. We then assume that the system undergoes a
phenomenological quench and the order parameter ($ <\sigma>  $) then rolls
toward the bottom of its potential. In this case the effective quark
mass is then {\em time dependent} which induces a time 
dependence in the triangle diagram and hence in the effective anomalous
electromagnetic coupling. Furthermore as the $ <\sigma> $ evolves,
long-wavelength pion fluctuations are generated that feed in to the
triangle diagram as well. We obtain the quark propagators during the
rolling of the chiral condensate 
in a mean field approximation. The effective pseudoscalar-photon interactions
are then obtained by explicitly integrating out the quarks and computing the
triangle diagram in real-time and a local limit is extracted for quasistatic
classical electromagnetic fields. Using this systematic formulation we then
derive the equation of motion for the neutral pion field after including both
(a) self-consistent mean-field effects and 
(b) effective nonequilibrium anomalous
couplings to the quasi-static semiclassical electromagnetic fields which are
of relevance in peripheral heavy-ion collisions. 
We then argue that unstable long-wavelength pion fluctuations can enhance the
isospin breaking
electromagnetic effects via the nonequilibrium anomalous coupling.

In an Appendix we also show that the anomaly equation and the axial Ward
identity are not modified by the medium in or out of equilibrium. Thus our
study shows that the Ward identity does {not} completely determine
the vertex in a medium and  more importantly, the latter is
sensitive to low energy physics and is therefore model-dependent.

In section II we obtain the effective action in real time in equilibrium,
analyze the infrared divergences and obtain the HTL corrections
to the self-energy. We analyze the different processes that lead to a width for
the fermionic quasiparticles and obtain a quantitative estimate for the local
limit of the effective pseudoscalar-photon vertex. [By local limit we
mean taking the external momenta to zero]. 

Section III is devoted to the nonequilibrium case. We consider a `quenched'
chiral phase transition in which the expectation value of the chiral order
parameter begins with small amplitude and rolls down its potential hill
towards the minimum equilibrium state. The nonequilibrium effective action for
pions and photons is obtained in the mean-field approximation. Finally we
obtain the effective nonequilibrium vertex for the neutral pion in the limit
of quasistatic classical electromagnetic fields, thus obtaining the effective
equation of motion for the neutral pion including its anomalous coupling.

Section IV summarizes our results. Appendix A deals with obtaining the
anomaly equation in real time both in and out of equilibrium, while
Appendix B provides the technical details of the solution of the Dirac equation
in a time dependent scalar background.

\section{$\pi\gamma\gamma$ vertex in equilibrium at finite temperature}

The purpose of this section is to provide a detailed analysis within real-time,
finite temperature field theory, of the effective anomalous coupling between
pseudoscalars and photons in a simple model that incorporates all of the
essential physics. We work within a Gell-Mann L\'evy type model containing a
single fermion flavour $\psi$ charged under $U(1)_{em}$, with Yukawa couplings
to a scalar $\sigma$ and a pseudoscalar $\pi$. As it will become clear below
the results of our analysis generalize straightforwardly to the $U(2)\otimes
U(2)$ constituent quark model with the appropriate inclusion of the isospin
group structure in the vertices. The Lagrangian density for the model under
consideration is given by


\begin{eqnarray}
{\cal L}=&&i\bar{\psi}(\partial\!\!\!\slash-ieA\!\!\!\slash)\psi
-2g\; \bar\psi(\sigma+i\pi\gamma_5)\psi
-\frac{1}{4}F_{\mu\nu}F^{\mu\nu}+{\cal L}_{\pi}.\label{sigmalag}
\end{eqnarray}

${\cal L}_{\pi}$
determines the dynamics of the $\pi$ and $\sigma$ fields and is  
a $\phi^4$ scalar theory given by, 
\begin{eqnarray}
&&{\cal L}_{\pi}=
\frac{1}{2}(\partial_{\mu}\sigma)^2 + \frac{1}{2}
(\partial_{\mu}\pi)^2 -{\lambda}
(\sigma^2+\pi^2-f_a)^2+c\sigma
\end{eqnarray}

The vacuum constituent quark mass $\approx 300\ \mbox{MeV} \approx 2g f_{\pi}$
with the vacuum value of $f_{\pi} = 93\ \mbox{MeV}$ leads to $g \approx 1.5 > e
\approx 0.3$. Although the large phenomenological value of the Yukawa coupling
would preclude a perturbative expansion, we will {\em assume} $1>>g>>e$ so as
to be able to provide a {\em quantitative} assessment of the in-medium effects at
least within a perturbative expansion. While the assumption $g<<1$ is necessary
if we want to provide a quantitative understanding based on perturbation
theory, the condition $e<<g$ is imposed only to simplify the calculations and
can be easily relaxed without affecting the qualitative validity of the results
in what follows.

A small linear term $ c\,\sigma $ ensures that in equilibrium isospin
symmetry is broken along the $ \sigma $ direction. At zero temperature
$\langle\sigma\rangle\approx f_{\pi}$ and the explicit symmetry breaking term
gives the pions their mass. The strict chiral limit refers to $
c\equiv 0 $.  At
finite temperatures $ T\neq 0 $, the expectation value of $ \sigma $ acquires
temperature dependence and we split the quantum field into a c-number
expectation value and quantum fluctuations:
\begin{eqnarray}
\sigma(x)=\sigma_0(T)+\chi(x)\;;\;\;\;\langle\sigma(x)\rangle=\sigma_0(T),
\end{eqnarray}
Here the expectation value of the Heisenberg operators is evaluated in the
thermal density matrix. The fermion thus gets a temperature dependent
tree-level mass,
\begin{eqnarray}
m_q(T)=2 g\; \sigma_0(T).
\end{eqnarray}

Our subsequent analysis will be valid in the regime where $
T>>g\;\sigma_0(T) $. In
particular this condition will be satisfied near the critical temperature where
in the strict chiral limit $c=0$, $\sigma_0(T)$ vanishes as
$ \sigma_0(0)\;\sqrt{1-\frac{T}{T_c}} \; ; \; T \rightarrow T_c^- $.

\subsection{The real time effective action:}

Within the simple constituent quark model introduced above, we now want to
study the effective theory of pions interacting with photons at a finite
temperature $T\neq 0$. Moreover, we would like to be able to study the {\em
dynamics} of such interactions. {\em Static} phenomena can  be studied
using the imaginary time formulation of finite temperature field theory.

The proper way to study {\em dynamical} phenomena in quantum field theory is
via the {\bf real time} evolution of an initially prepared density
matrix. This is implemented in the path integral formulation by use of
the Closed Time Path (CTP) formalism of Keldysh and Schwinger
\cite{ctp,keldysh}. This formalism allows us to compute 
the time evolution of expectation values of Heisenberg operators and can be
applied both to real time equilibrium calculations as well as nonequilibrium
ones. In this formulation the action is defined on a contour that runs forward
and backward in time as befits the time evolution of an initially prepared
density matrix. Field operators on the forward and backward branches are
denoted with $+$ and $-$ superscripts respectively and are independent field
variables. For a description of this formulation within the setting of problems
akin to those studied here see
\cite{ctp,keldysh,boyadcc,boyareheat,largeN}. 

As we have stressed in the introduction our goal is to derive an effective
description for the pions and photons which really means obtaining the real
time effective action by integrating out the quarks including the
anomalous pion-photon interactions. We remark that, for our purposes the {\em
decay amplitude} for $\pi\rightarrow\gamma\gamma$ as defined in \cite{gelis}
for 
example, is a more restricted quantity than the full real-time effective
action. The former is given by a particular cut of the self-energy graph
(Fig. 2) for the $\pi$. The couplings in the effective action also contain
discontinuities arising 
from the on-shell propagation of intermediate fermions in the thermal
bath. Cross-couplings between fields living on the forward and backward
contours will also be induced.
{\em All} of these terms in the effective
action will be important for understanding the effective pion-photon
interactions   in the thermal bath; for
example the full real-time equation of motion for pion condensates will
necessarily receive contributions from all these different terms.

Although the effective action in Euclidean time has been derived previously, we
obtain here the real time effective action, which we believe is a novel
approach. 
The utility of the real time effective action lies in the fact that it 
allows us to obtain the real-time, retarded equations of motion for expectation
values even {\em out of equilibrium}. As emphasized above the focus of this
article is to provide a real-time description of non-equilibrium phenomena.
Since the real time effective action within this setting has not yet received
attention, we begin by studying the equilibrium situation first to compare with
previous results in the literature and then continue with the focus of our
program, the non-equilibrium 
aspects. 

In order to obtain the aforementioned effective couplings we integrate out the
fermions to one-loop to get the (CTP) effective action for the
pseudoscalars and 
the photons up to this order. Formally this is a rather straightforward
procedure, the full real-time effective action being given by:
\begin{eqnarray}
&&\exp{iS_{eff}[A_\mu^{\pm}, \pi^\pm]}=\nonumber\\\nonumber\\ 
&&=\int [{\cal D}\psi^{\pm}][{\cal D}\bar\psi^{\pm}]
\exp[i\int
d^3x\;dt\{i\bar\psi^+(i\partial\!\!\!\slash-2g\sigma_0)\psi^+  
-(+\longrightarrow-)\}]\times\\\nonumber\\\nonumber
&&\times\exp[i\int d^4x\{e\bar\psi^+\gamma^\mu A_\mu^+\psi^+
-2g\bar\psi^+(\chi^++i\pi\gamma_5)\psi^+-(+\longrightarrow-)\}].
\end{eqnarray}
The one-loop contribution is obtained by simply expanding the interaction part
$\exp[iS_{int}]$ to $O(e^2g)$ and integrating out the
fermions yielding the following expression for the $\pi\gamma\gamma$
interactions: 
\begin{eqnarray}
&&S_{\pi\gamma\gamma}= -2ge^2\int
d^4x\;d^4x_1\;d^4x_2\times\nonumber\\\nonumber\\\label{seff}
&&\times\left\{\pi^{+}(x)A_\mu^+(x_1)A_\nu^+(x_2)\;\;
\mbox{Tr}[S^{++}(x_2,x)\gamma_5 S^{++}(x,x_1)\gamma^\mu
S^{++}(x_1,x_2)\gamma^\nu] 
\right.\\\nonumber\\\nonumber 
&&\left.-\pi^{+}(x)A_\mu^+(x_1)A_\nu^-(x_2)\;\;
\mbox{Tr}[S^{-+}(x_2,x)\gamma_5S^{++}(x,x_1)\gamma^\mu
S^{+-}(x_1,x_2)\gamma^\nu] 
\right.\\ \nonumber\\\nonumber
&&\left.-\pi^+(x)A_\mu^-(x_1)A_\nu^+(x_2)\;\;
\mbox{Tr}[S^{++}(x_2,x)\gamma_5S^{+-}(x,x_1)\gamma^\mu
S^{-+}(x_1,x_2)\gamma^\nu] 
\right.\\\nonumber\\\nonumber
&&\left.+\pi^{+}(x)A_\mu^-(x_1)A_\nu^-(x_2)\;\;
\mbox{Tr}[S^{-+}(x_2,x)\gamma_5S^{+-}(x,x_2)\gamma^\nu
S^{--}(x_2,x_1)\gamma^\mu] -(\pi^+\rightarrow\pi^-)\right\} \; .
\end{eqnarray} 
where $S^{-+}, S^{+-}, S^{++}$ and $S^{--}$ are the real time fermionic
propagators given by: 
\begin{eqnarray}
&&S^{\pm\pm}(x,x^\prime)=-i\, 
\langle\psi^\pm(x){\bar\psi}^\pm(x^\prime)\rangle;\\\nonumber\\
&&S^{-+}(x,x^\prime)=-i\, \langle\psi^-(x){\bar\psi}^+(x^\prime)\rangle
;\;\;\;\;\; 
S^{+-}(x,x^\prime)=i\, \langle{\bar\psi}^+(x^\prime)\psi^-(x)\rangle
;\\\nonumber\\
&&S^{++(--)}(x,x^\prime)=-S^{-+(+-)}(x,x^\prime)\; \Theta(x_0-x_0^\prime)-
S^{+-(-+)}(x-x^\prime)\;\Theta(x_0^\prime-x_0)\label{fermprops}.
\end{eqnarray} 

Note that the interactions have induced cross-couplings between fields on
different branches of the CTP contour. This means that answers to physical
questions such as how the pion couples to classical electromagnetic sources
will be affected by the interplay between all these cross-couplings in the
effective action. In particular the equation of motion for the expectation
value of the pion field will require {\em all} of these
contributions\cite{boyadcc,boyareheat}.  Evaluation of the various terms in
the effective action is more convenient in momentum space and in the
equilibrium case we can introduce the Fourier-transformed fields and
propagators:
\begin{eqnarray}
&&\tilde{\pi}(k)=\int d^4x\;e^{ik\cdot x}\; \pi(x)\;,\\\nonumber\\
&&\tilde{A}_\mu(k)=\int d^4x\;e^{ik\cdot x}\; A_\mu(x)\;,\\\nonumber\\
&&S(k)=\int d^4x \;e^{ik\cdot (x-x^\prime)}\; S(x,x^\prime)\;.
\end{eqnarray}
The effective action in momentum space then takes the following form
\begin{eqnarray}
&&S_{\pi\gamma\gamma}=-2e^2g\int 
\frac{d^4p_1}{(2\pi)^4}\frac{d^4p_2}{(2\pi)^4}\times\label{seffk}\\\nonumber
&&\times\left\{\tilde\pi^{+}(-p_1-p_2)\tilde
A_\mu^+(p_1) 
\tilde{A}_\nu^+(p_2)
\int\frac{d^4k}{(2\pi)^4}
\mbox{Tr}[S^{++}(k-p_1)\gamma_5S^{++}(k+p_2)\gamma^\mu S^{++}(k)\gamma^\nu]
+\right.\\\nonumber\\\nonumber
&&\left.-\tilde\pi^{+}(-p_1-p_2)\tilde
A_\mu^+(p_1)\tilde{A}_\nu^-(p_2)
\int\frac{d^4k}{(2\pi)^4}
\mbox{Tr}[S^{-+}(k-p_1)\gamma_5S^{++}(k+p_2)\gamma^\mu S^{+-}(k)\gamma^\nu]
+\right.\\\nonumber\\\nonumber
&&\left.-\tilde\pi^{+}(-p_1-p_2)\tilde
A_\mu^-(p_1)\tilde{A}_\nu^+(p_2)
\int\frac{d^4k}{(2\pi)^4}
\mbox{Tr}[S^{++}(k-p_1)\gamma_5S^{+-}(k+p_2)\gamma^\mu S^{-+}(k)\gamma^\nu]
+\right.\\\nonumber\\\nonumber
&&\left.+\tilde\pi^{+}(-p_1-p_2)\tilde
A_\mu^-(p_1)\tilde{A}_\nu^-(p_2)
\int\frac{d^4k}{(2\pi)^4}
\mbox{Tr}[S^{-+}(k-p_1)\gamma_5S^{+-}(k+p_2)\gamma^\mu S^{--}(k)\gamma^\nu]
\right\}.\\\nonumber
\end{eqnarray}

>From the definitions of the real time propagators (\ref{fermprops}) it is clear
that the four propagators satisfy the identity:
\begin{eqnarray}
S^{++}+S^{+-}+S^{-+}+S^{--}=0,
\end{eqnarray}
which of course continues to hold for the Fourier-transformed
propagators as well. 
So we can rewrite them in terms of three independent functions in the Keldysh
\cite{keldysh} notation,
\begin{eqnarray}
S_R(k)=S^{++}(k)+S^{+-}(k),\label{keldysh1}\\\nonumber\\
S_A(k)=S^{++}(k)+S^{-+}(k),\label{keldysh2}\\\nonumber\\
S_F(k)=S^{++}(k)+S^{--}(k).\label{keldysh3}
\end{eqnarray}  
Here $S_R$, $S_A$ are the retarded and advanced propagators, while $S_F$ is a
symmetric combination which is closely related to the spectral density in
thermal equilibrium:
\begin{eqnarray}
&&S_F(k)
=[1-2n_f(|k_0|)]\mbox{sgn}(k_0)[S_R(k)-S_A(k)]\\\nonumber\\
&&n_f(|k_0|)=\frac{1}{e^{\beta|k_0|}+1}.
\end{eqnarray}
This relationship can be shown to arise via the Lehmann spectral representation
for states in thermal equilibrium and it  does not hold out-of-equilibrium.
All these definitions and relations are exact and continue to hold for the full
field theory in thermal equilibrium.

\subsection{The `bare' $\pi\gamma\gamma$ vertex:}

The anomalous couplings in the CTP effective action are to be thought of 
in terms of the triangle diagram in Fig. 1 with insertions of
external fields at the vertices, carrying all possible combinations of  $+$ and
$-$ labels on them, corresponding to the
relevant portions of the CTP contour \cite{boyadcc,boyareheat,largeN}. 

Since we
want to study the low-energy effective couplings, we must first clearly
identify the scales involved in the problem.  The external photons
have 4-momenta $ p_1,p_2 $
(see Fig. 1 for example) while the pion has momentum $ -p_1-p_2 $ by momentum
conservation. 
Henceforth we will collectively denote the momenta of the
external fields by $ P_{\mbox{\small ext}} $. By low-energy vertices we will
always mean effective interactions where $ P_{\mbox{\small ext}} << T $, and
since we want to remain close to the chiral phase transition, we
choose  $ m_q<<T $ as well. 
From our subsequent analysis it will
become clear that other scales such as 
$ g \, T, g^2 \,T $, originating dynamically from the thermal
contributions to the fermion self-energy will also be extremely important.
(Recall, in this connection our initial assumption that $
g<<1 $ so
that there is indeed a possible hierarchy of scales $ T,g \,T.. $, etc.)

\vspace{0.5in}
\centerline{\epsfbox{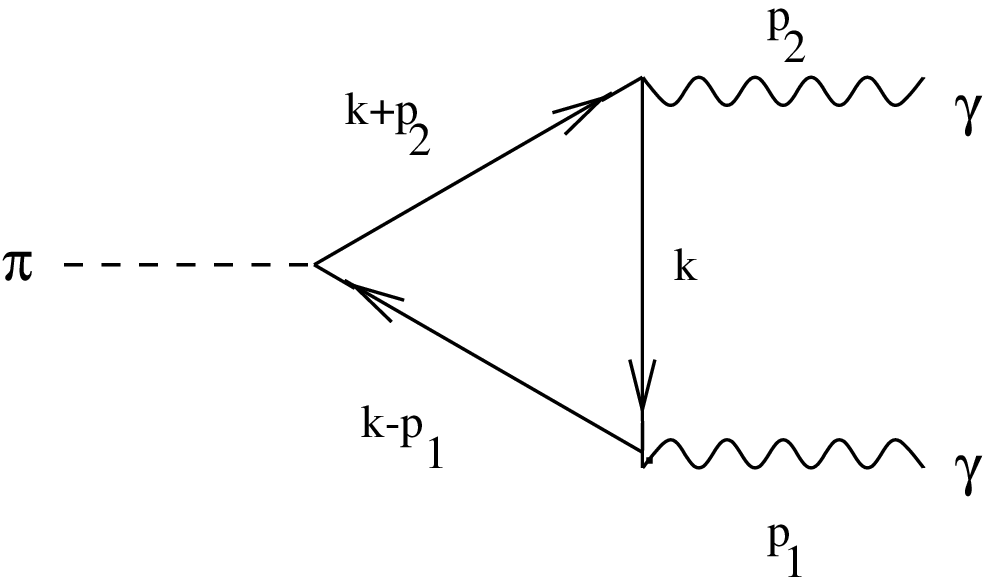}}
\vspace{0.5 in}

{FIG. 1. {\small{The triangle diagram at finite temperature with bare 
fermion propagators in the internal lines.}}}\\

First consider the regime $g \,T<<m_q<<T $ and $ {P}_{\mbox{\small
ext}}<<m_q $.  
In order to provide motivation for the other scales that will be important at
finite $T$ let us first calculate the `bare' $ \pi\gamma\gamma $
vertex, where 
by `bare' we are referring to an evaluation of the terms in the effective
action in Eq. ({\ref{seff}}) with {\em undressed} finite temperature
propagators for 
the fermions in the loop (Fig. 2). At finite temperature the free fermion
Green's functions in the real-time approach are:
\begin{eqnarray}
&&S^{\pm\pm}(k)=(\not k+m_q)\left[\pm\frac{1}{k^2-m^2_q\pm i\epsilon}+
2\pi i\;n_f(\omega_k)
\delta(k^2-m^2_q)\right]\\\nonumber\\
&&S^{\mp\pm}(k)=2\pi i\;
(\not k+m_q) \delta(k^2-m^2_q)\left[\Theta(\pm k_0)-n_f(\omega_k)\right]\\\nonumber\\
&&n_f(|k_0|)=\frac{1}{e^{\beta k_0}-1};\;\;\;\omega_k=\sqrt{k^2+m_q^2}.
\end{eqnarray}

The local form of the low energy vertex $\pi^+A^+A^+$ (from Eq. (\ref{seffk}))
is obtained by setting $P_{\mbox{\small ext}}=0 $ in the computation of the
triangle diagram, (as was done in the imaginary time formulation in
\cite {rob,robtyt,baier}). 
However, in the real time
formulation, setting the external momenta to zero inside the quark loop leads
to a {\em divergent} result:
\begin{eqnarray}
S_{\pi\gamma\gamma}^{+++}=&&8m_q\;e^2g\int 
\frac{d^4p_1}{(2\pi)^4}\frac{d^4p_2}{(2\pi)^4}\times
\left\{i\;\epsilon^{\alpha\beta\mu\nu}\;p_{1\alpha}p_{2\beta}\;
\tilde\pi^{+}(-p_1-p_2)
\tilde A_\mu^+(p_1) \tilde{A}_\nu^+(p_2)\right.\label{badvertex}
\\\nonumber\\\nonumber
&&\left.\int\frac{d^4k}{(2\pi)^4}
\left[\frac{1}{k^2-m^2_q+i\epsilon}+
2\pi i\;n_f(\omega_k)\;
\delta(k^2-m^2_q)\right]^3\right\}
\end{eqnarray}

This integral has two different sources of infrared divergences. In the chiral
limit $m_q \rightarrow 0$ the temperature independent term becomes sensitive to
the infrared modes. At finite temperature this divergence must be cured by
resumming the contributions from HTL.  On the other hand the
temperature dependent terms lead to divergences $\sim T^2/\epsilon^2$. This is
a 
pinch singularity that originates from the propagation of on-shell fermionic
intermediate states present in the bath. At tree level, these quarks in the
plasma propagate without damping over arbitrarily long times. However, in an
interacting plasma the quarks will acquire a width $\Gamma_k$ and will only
propagate during time scales $\approx 1/\Gamma_k$. Including this effect in the
fermion propagator will cut off these pinch singularities.

This naive calculation is obviously flawed but clearly indicates that using
bare propagators and setting ${P}_{\mbox{\small ext}}$ to zero is inconsistent
with the high temperature limit. If we imagine taking $P_{\mbox{\small ext}}$
to zero continuously, at some value the correlator will become sensitive to
thermal corrections to the self-energy. The real part of the thermal
self-energy should play the role of $m_q$ in the loop while the on-shell
thermal width should replace the Feynman parameter $\epsilon$.%
\footnote{One of us (S.P.K.) would like to thank S. Jeon and L. Yaffe for
pointing this out.}

Before proceeding to a more consistent description using resummed propagators,
we seek to establish contact with the results of \cite{rob} and
\cite{gelis} using the real time effective
action obtained above. We do the calculation by keeping the external momenta
non-zero, and then take the zero momentum limit smoothly at the end of the
calculation.

Using bare fermion propagators and defining
\begin{eqnarray}
&&\omega=\sqrt{k^2+m_q^2}\;;\;\;
\omega_1=\sqrt{(\vec k -\vec p_1)^2+m_q^2}\;;\;\;
\omega_2=\sqrt{(\vec k +\vec p_2)^2+m_q^2}\;;\\\nonumber\\
&&n=n_f(\omega)\;;\;\;n_1=n_f(\omega_1)\;;\;\;n_2=n_f(\omega_2)\;,\\\nonumber
\end{eqnarray}
we find that the $\pi^+A_\mu^+A_\nu^+$ coupling in the effective
action is given by:
\begin{eqnarray}
\nonumber\\
&&S_{\pi\gamma\gamma}^{+++}=-m_q\;e^2g\int 
\frac{d^4p_1}{(2\pi)^4}\frac{d^4p_2}{(2\pi)^4}\times
\epsilon^{\alpha\beta\mu\nu}\;p_{1\alpha} \; p_{2\beta}\;
\tilde\pi^{+}(-p_1-p_2) \;\tilde A_\mu^+(p_1) \;
\tilde{A}_\nu^+(p_2)\times\label{plplpl} 
\\\nonumber\\\nonumber
&&\int\frac{d^3k}{(2\pi)^3}\frac{1}{\omega\omega_1\omega_2}
\left\{(1-n_1)(1-n_2)n
\left[\frac{1}{(\omega_1-\omega-p_1^0-i\epsilon)
(\omega_2-\omega+p_2^0-i\epsilon)}+\right.\right.\\\nonumber\\\nonumber
&&\left.\left.
\frac{1}{(\omega_2-\omega-p_2^0-i\epsilon)
(\omega_1-\omega+p_1^0-i\epsilon)}+\frac{1}{(\omega_2-\omega-p_2^0-i\epsilon)
(\omega_2+\omega_1-p_2^0-p_1^0-i\epsilon)}+
\right.\right.\\\nonumber\\\nonumber
&&\left.\left.
\frac{1}{(\omega_1-\omega-p_1^0-i\epsilon)
(\omega_1+\omega_2-p_1^0-p_2^0-i\epsilon)}+
\frac{1}{(\omega_1-\omega+p_1^0-i\epsilon) 
(\omega_1+\omega_2+p_1^0+p_2^0-i\epsilon)}+
\right.\right.\\\nonumber\\\nonumber
&&\left.\left.
\frac{1}{(\omega_2-\omega+p_2^0-i\epsilon)
(\omega_1+\omega_2+p_1^0+p_2^0-i\epsilon)}\right]
-n_1n_2(1-n)\left({\mbox {c.c.}}\right)+\right.\\\nonumber\\\nonumber
&&\left.n_1(1-n_2)(1-n)
\left[\frac{1}{(\omega-\omega_1+p_1^0-i\epsilon)
(\omega+\omega_2-p_2^0-i\epsilon)}+\right.\right.\\\nonumber\\\nonumber
&&\left.\left.
\frac{1}{(\omega+\omega_2+p_2^0-i\epsilon)
(\omega-\omega_1-p_1^0-i\epsilon)}+\frac{1}{(\omega_2+\omega+p_2^0-i\epsilon)
(\omega_2-\omega_1+p_1^0+p_2^0-i\epsilon)}+
\right.\right.\\\nonumber\\\nonumber
&&\left.\left.
\frac{1}{(\omega-\omega_1+p_1^0-i\epsilon)
(\omega_2-\omega_1+p_1^0+p_2^0-i\epsilon)}+\frac{1}{(\omega-\omega_1-p_1^0-i\epsilon)
(\omega_2-\omega_1-p_1^0-p_2^0-i\epsilon)}+
\right.\right.\\\nonumber\\\nonumber
&&\left.\left.
\frac{1}{(\omega_2+\omega-p_2^0-i\epsilon)
(\omega_2-\omega_1-p_1^0-p_2^0-i\epsilon)}\right]
-(1-n_1)n_2n\left[{\mbox {c.c.}}\right]+\right.\\\nonumber\\\nonumber
&&\left.(1-n_1)n_2(1-n)
\left[\frac{1}{(\omega_1+\omega-p_1^0-i\epsilon)
(\omega-\omega_2+p_2^0-i\epsilon)}+\right.\right.\\\nonumber\\\nonumber
&&\left.\left.
\frac{1}{(\omega-\omega_2-p_2^0-i\epsilon)
(\omega+\omega_1+p_1^0-i\epsilon)}+\frac{1}{(\omega-\omega_2-p_2^0-i\epsilon)
(\omega_1-\omega_2-p_1^0-p_2^0-i\epsilon)}+
\right.\right.\\\nonumber\\\nonumber
&&\left.\left.
\frac{1}{(\omega+\omega_1-p_1^0-i\epsilon)
(\omega_1-\omega_2-p_1^0-p_2^0-i\epsilon)}+\frac{1}{(\omega+\omega_1+p_1^0-i\epsilon)
(\omega_1-\omega_2+p_1^0+p_2^0-i\epsilon)}+
\right.\right.\\\nonumber\\\nonumber
&&\left.\left.
\frac{1}{(\omega-\omega_2+p_2^0-i\epsilon)
(\omega_1-\omega_2+p_1^0+p_2^0-i\epsilon)}\right]
-n_1(1-n_2)n\left[{\mbox {c.c.}}\right]+\right.\\\nonumber\\\nonumber
&&\left.n_1n_2n
\left[\frac{1}{(\omega+\omega_1+p_1^0+i\epsilon)
(\omega+\omega_2-p_2^0+i\epsilon)}+\right.\right.\\\nonumber\\\nonumber
&&\left.\left.
\frac{1}{(\omega+\omega_2+p_2^0+i\epsilon)
(\omega+\omega_1-p_1^0+i\epsilon)}+\frac{1}{(\omega+\omega_2+p_2^0+i\epsilon)
(\omega_1+\omega_2+p_1^0+p_2^0+i\epsilon)}+
\right.\right.\\\nonumber\\\nonumber
&&\left.\left.
\frac{1}{(\omega+\omega_1+p_1^0+i\epsilon)
(\omega_1+\omega_2+p_1^0+p_2^0+i\epsilon)}+\frac{1}{(\omega+\omega_1-p_1^0+i\epsilon)
(\omega_1+\omega_2-p_1^0-p_2^0+i\epsilon)}+
\right.\right.\\\nonumber\\\nonumber
&&\left.\left.
\frac{1}{(\omega+\omega_2-p_2^0+i\epsilon)
(\omega_1+\omega_2-p_1^0-p_2^0+i\epsilon)}\right]
-(1-n_1)(1-n_2)(1-n)\left[{\mbox {c.c.}}\right]\right\}\\\nonumber
\end{eqnarray}

The above interaction is just one of the couplings that appears in the real
time effective 
action (others are $\pi^+A_{\mu}^+A_\nu^-,\pi^+A_{\mu}^-A_\nu^-$, etc.). An
important point to note is that aside from the direct interactions with 
photons it also contains discontinuities from the on-shell propagation of
fermionic 
intermediate 
states that appear in the bath and couple to the external electromagnetic
field. These are encoded in the  
$i\epsilon$ terms of the energy denominators and give rise to pinch
singularities when external momenta are set to zero as  previously discussed.  

We now make contact with the results of \cite{rob,gelis,gnayak} by calculating
the decay amplitude $\Gamma(\pi\rightarrow\gamma\gamma)$ which can be obtained
from 
the above 
expression by simply {\em ignoring} the discontinuities in this effective
vertex. The resulting expression can be interpreted as a particular cut  
of the 3-loop self-energy graph \cite{gelis} (see Fig. 2) for the $\pi$ where
the photons appear as final states. 

In the limit of $P_{\mbox{\small ext}}<<T$ 
and $P_{\mbox{\small ext}}<<m_q$, the distribution functions
and the energy-denominators can be expanded in powers of $P_{\mbox{\small
ext}}/T$ and $P_{\mbox{\small ext}}/m_q$ respectively to yield the amplitude
(or the 
imaginary part of the retarded self-energy) for
$\pi\rightarrow\gamma\gamma$. After long-winded algebraic manipulations we
obtain the following form for the vertex in the zero-momentum limit:

\vspace{0.5in}
\centerline{\epsfbox{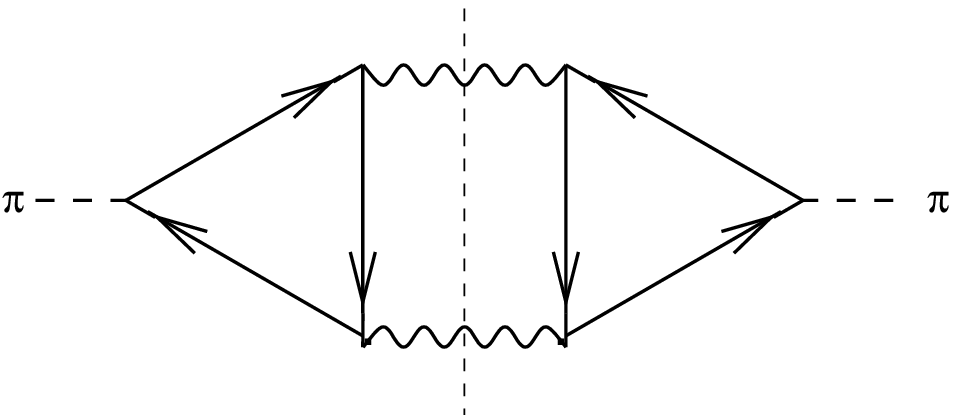}}
\vspace{0.5 in}

{FIG. 2: {\small{The 3-loop self-energy cut to yield the required decay
amplitude in \cite {gelis}.}}}\\ 

\begin{eqnarray}
&&\Gamma(\pi\rightarrow\gamma\gamma)=
m_q\;e^2g\;\int\frac{d^3k}{(2\pi)^3}\frac{1}{\omega^3}\times\label{allamplitude}
\\\nonumber\\\nonumber
&&\left\{\frac{3(1-2n)}{2\omega^2}+
\frac{dn(\omega)}{d\omega}\frac{\vec k\cdot\vec p_1}{2\omega}
\left[\frac{1}{\tilde k\cdot p_1}-\frac{1}{k\cdot p_1}+\frac{1}{\tilde
k\cdot(p_1+p_2)} 
-\frac{1}{k\cdot(p_1+p_2)}
+\right.\right.\\\nonumber\\\nonumber
&&\left.\left.+
\frac{2\vec k \cdot\vec p_1}{k\cdot p_1 k\cdot (p_1+p_2)}
+\frac{2\vec k \cdot\vec p_1}{\tilde k\cdot p_1\tilde k\cdot (p_1+p_2)}-
\frac{2\vec k \cdot\vec p_2}{k\cdot p_2 k\cdot (p_1+p_2)}-
\frac{2\vec k \cdot\vec p_2}{\tilde k\cdot p_2\tilde k\cdot (p_1+p_2)}
\right]\right.\\\nonumber\\\nonumber
&&\left.
+\frac{dn(\omega)}{d\omega}
\frac{\omega^2 {\vec p_1}^2-(\vec k\cdot\vec p_1)^2}{2\omega}\left[
\frac{\vec k\cdot\vec p_1}{(\tilde k\cdot p_1)^2\tilde k\cdot (p_1+p_2)}
-\frac{\vec k\cdot\vec p_1}{(k\cdot p_1)^2k\cdot (p_1+p_2)}
\right.\right.\\\nonumber\\\nonumber
&&\left.\left.
+\frac{\vec k\cdot\vec p_1}{\tilde k\cdot p_1(\tilde k\cdot (p_1+p_2))^2}
-\frac{\vec k\cdot\vec p_1}{k\cdot p_1 (k\cdot (p_1+p_2))^2}
-\frac{\vec k\cdot\vec p_2}{\tilde k\cdot p_2(\tilde k\cdot (p_1+p_2))^2}
\right.\right.\\\nonumber\\\nonumber
&&\left.\left.
+\frac{\vec k\cdot\vec p_2}{k\cdot p_2(k\cdot (p_1+p_2))^2}
-\frac{1}{k\cdot p_1 k\cdot (p_1+p_2)}
-\frac{1}{\tilde k\cdot p_1\tilde k\cdot (p_1+p_2)}
\right]\right.\\\nonumber\\\nonumber
&&\left.
-\frac{1}{2}\frac{d^2n(\omega)}{d\omega^2}(\vec k\cdot\vec p_1)^2
\left[\frac{1}{k\cdot p_1k\cdot(p_1+p_2)}+
\frac{1}{\tilde k\cdot p_1\tilde k\cdot(p_1+p_2)}
\right]+(p_1\leftrightarrow p_2)\right\}\\\nonumber\\
&&\mbox{where}\;\;\omega= \sqrt{\vec{k}^2+m^2_q}\;\; , \;\; \tilde{k} =
(\omega, -\vec{k})\; .  
\end{eqnarray}

The symmetrization with respect to $p_1$ and $p_2$ applies only to the
momentum-dependent terms. This is one of the important results of this paper. 
Using this expression one can calculate the required amplitude for different
external kinematical configurations. 
For spacelike photons $(p_1)^0=(p_2)^0=0$ 
and in the limit $ m_q/T\rightarrow 0 $ the expression turns out to be
finite and is given by: 

\begin{eqnarray}
\Gamma(\pi\rightarrow\gamma\gamma)=
m_q\;e^2g\;\int\frac{d^3k}{(2\pi)^3}\;\frac{1}{k^3}
\left\{\frac{3}{2k^2}[1-2n(k)]+
{3\over k}\;\frac{dn(k)}{dk} -\frac{d^2n(k)}{dk^2}\right\} \; .
\label{spacelike}\end{eqnarray}

This expression agrees with the result of \cite{gelis} and after careful
integrations by parts can be easily shown to yield the zeta-function dependent
result of the Euclidean calculation. It behaves as $ \sim m_q \; e^2
\,g/T^2 $ in
agreement with \cite{rob,robtyt}.

In the case of pion decay at rest to two back-to-back on-shell photons 
we find
after straightforward  substitution

\begin{eqnarray}
&&\Gamma(\pi\rightarrow\gamma\gamma)=\\\nonumber
&&m_q\;e^2g\;\int\frac{d^3k}{(2\pi)^3}\;\frac{1}{\omega^3}
\left\{\frac{3}{2\omega^2}[1-2n(\omega)]-
{1\over \omega}\;\frac{dn(\omega)}{d\omega} -
\frac{d^2n(\omega)}{d\omega^2}\frac{(\vec k\cdot \vec p)^2}
{\omega^2p^2-(\vec k \cdot p)^2}
\right\} \; .
\label{lightlike}\end{eqnarray}

This coincides with the expression found in \cite{gelis} and can be shown to
be equivalent to the results obtained in \cite{gnayak}.
In this case and also for other generic kinematical configurations
Eq.(\ref{allamplitude}) in fact behaves as $ \sim e^2 \,g/T $. 
These results are, however, valid only in the regime where
$ T>>m_q>>g \,T $ so that we can ignore the hard thermal loops.

While this calculation seems to indicate that self-energy corrections will not
play a role in the amplitude, the terms in the effective
action also involve additional contributions from thermal discontinuities. 
These are the terms that yield pinch singularities in the zero
momentum limit, as can be easily checked by setting the external momenta {\em
equal} to zero in the exact expression for $ S^{+++}_{\pi\gamma\gamma}$
Eq. (\ref{plplpl}). 

The result obtained for the finite temperature vertex in the imaginary time
formulation in \cite{rob,robtyt,baier} cannot be extrapolated to the real time
domain. In \cite{rob,robtyt} the calculation was performed by setting
the external 
Matsubara frequencies to zero. In order to obtain the real time behavior, the
external Matsubara frequencies must be kept non-zero in the evaluation
of the triangle diagram and the resulting expression will therefore be
a function of the external frequencies $\omega_n$ defined at the
discrete points along the imaginary axis in 
the frequency plane. The real-time vertex is then obtained by the continuation
$\omega_n \rightarrow i(\omega \pm i\epsilon)$; the imaginary parts resulting
from the $\pm i \epsilon$ will reveal the different discontinuities and
therefore the ensuing pinch singularities in the limit of soft external
four-momentum.

\subsection{The dressed vertex with HTL resummation:}

From the above expressions it is clear that the amplitude is sensitive to
momenta $\sim m_q$ and as $m_q/T$ is taken to zero we have to include
self-energy corrections to the quark propagator. Furthermore we have seen that
the interaction terms in the effective action also include the discontinuities
from the quarks in the loop and these in turn will be sensitive to the thermal
width on-shell when the external momenta are sufficiently small (i.e. $P_{\mbox
{\small ext}}<<\Gamma$ where $\Gamma$ is the on-shell width for the fermionic
quasiparticles.).

\vspace{0.5in}
\centerline{\epsfbox{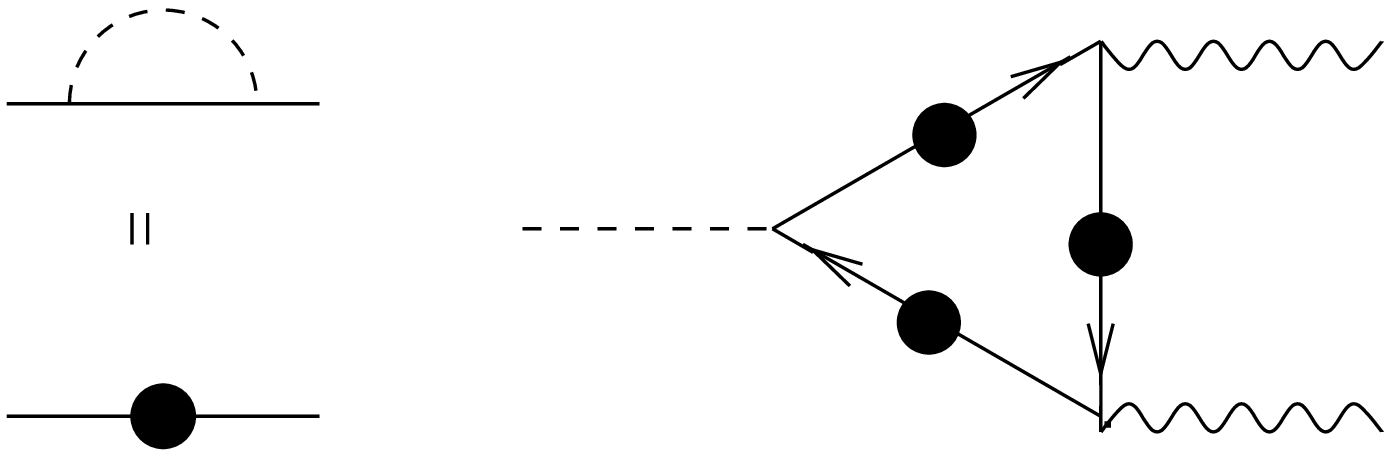}}
\vspace{0.5 in}

{FIG. 3 . {\small{The triangle diagram at finite temperature with resummed  
fermion propagators in the loops.}}}\\

In the high temperature, long-wavelength limit, i.e. when $ m_q\; ; \; P_{\mbox
{\small ext}}<<g \, T $ HTL resummed propagators must be introduced 
for the quarks in the loop (see Fig. 3). The theory obtains HTL corrections 
both from the gauge bosons and the scalar sector invoving the $ \sigma
$ and the $ \pi $.  However, since we have chosen $ e<<g $ we will ignore the
contributions from the photons to HTL. The $ \sigma $ and the $ \pi $
contribute equally to the thermal self-energy for the soft quarks. In
addition there are 
also the usual zero temperature logarithmic corrections to the fermion mass
$m_q$ which we shall ignore since these are not relevant for the
discussion. The retarded self-energy corrections from the scalars were
obtained in \cite{shang} and we summarize the main ingredients that
are relevant to our discussion here,  in the limit $ m_q=0 $. The one loop 
retarded self-energy for the quarks is given by $
\Sigma(k_0-i\epsilon,\vec k) = 
\Sigma^{\sigma}(k_0-i\epsilon,\vec k)+\Sigma^{\pi}(k_0-i\epsilon,\vec k)$ with 

\begin{equation}
\Sigma^a(k_0-i\epsilon,\vec k)=
i\; \gamma_0\;k_0 \;\tilde{\varepsilon}^{(0)}_{\vec{k}} (k_0-i\epsilon)+
\vec{\gamma}\cdot \vec{k}~~\tilde{\varepsilon}^{(1)}_{\vec{k}}(k_0-i\epsilon).\label{epsilons}
\end{equation} 
\noindent for $a= \sigma,\pi$. We define 
\begin{equation}
\left\{ \begin{array}{c}
\tilde{\varepsilon}^{(0)}_{\vec{k}}(k_0-i\epsilon)  \\
\tilde{\varepsilon}^{(1)}_{\vec{k}}(k_0-i\epsilon)  
\end{array} \right\} = 
 \int ds \frac{1}{(k_0-i\epsilon)^2+s^2} \left\{\begin{array}{c}
\rho^{(0)}_{\vec{k}}(s)  \\
s\; \rho^{(1)}_{\vec{k}}(s)   
\end{array}\right\}\label{dispersion}
\end{equation}
using the one-loop spectral densities  given  by the expressions:
\begin{eqnarray}
\nonumber\\
\rho^{(0)}_{\vec{k}}(s)&=& 4g^2 \int \frac{d^3 q}{(2\pi)^3
}\frac{\bar{\omega}_{q}}{
2 \omega_{k+q} \bar{\omega}_{q}} \times\nonumber\\\nonumber
&&\left[\delta(s-\omega_{k+q}-\bar{\omega}_{q})
(1+n_{k+q}-\bar{n}_{q})+ \delta(s-\omega_{k+q}+\bar{\omega}_{q})
(n_{k+q}+\bar{n}_{q}) \right]~,
\nonumber\\\nonumber\\\nonumber
\rho^{(1)}_{\vec{k}}(s)&=& 4g^2 \int \frac{d^3 q}{(2\pi)^3
} \frac{1}{2 \omega_{k+q} \bar{\omega}_{q}}\frac{\vec k \cdot \vec
q}{k^2}\times\nonumber\\ 
&&\left[\delta(s-\omega_{k+q}-\bar{\omega}_{q})
(1+n_{k+q}-\bar{n}_{q})-\delta(s-\omega_{k+q}+\bar{\omega}_{q})
(n_{k+q}+\bar{n}_{q})\right],\label{rhos}
\\\nonumber
\end{eqnarray}
where $n_{\vec k+\vec q}\; , \; \bar{n}_q$ are the Bose-Einstein and
Fermi-Dirac distribution functions and $\omega \; , \; \bar{\omega}$ are the
bosonic and fermionic frequencies respectively. The Bose-Einstein distributions
for the $\sigma$ and $\pi$ fields are functions of their respective masses.
We have also neglected the ultraviolet counterterms as
these are irrelevant for our
discussion. 

The terms proportional to $\delta(s-\omega_{k+q}-\bar{\omega}_{q})$
in the spectral densities correspond to the processes $q\leftrightarrow
\phi+q$, where $\phi$ is either a $\sigma$ or $\pi$ field, and
determine the usual two particle cut that survives in the zero temperature
limit. The terms proportional to $\delta(s-\omega_{k+q}+\bar{\omega}_{q})$
correspond to Landau damping and processes of the type $\phi\leftrightarrow
q+\bar{q}$, which can only occur in the medium. Near the critical temperature and
in the strict chiral limit, $m_{\sigma}, m_{\pi} \rightarrow 0$ as
$T\rightarrow T_c^-$ and $m_q \rightarrow 0$. Therefore, in this limit, 
the Landau damping terms lead to a HTL contribution to the quark propagator.

We extract the HTL contribution to the quark self-energy in the
strict chiral limit and near the critical temperature so that
$ T>>m_q,m_{\sigma},m_{\pi},P_{\mbox{ext}} $ in the usual
manner\cite{weldon,pis,kapusta,lebellac,blaizot} and find
\begin{eqnarray}
&&{\Sigma}_{HTL}(k_0-i\epsilon; \vec{k})= 
-\frac{M^2_{eff}}{k}\left\{\gamma_0\ln
\frac{k_0+k-i\epsilon}{k_0-k-i\epsilon}+\vec{\gamma}\cdot\hat{k}
\left[2-\frac{k_0}{k}\ln\frac{k_0+k-i\epsilon}{k_0-k-i\epsilon}
\right]\right\} \nonumber \\
&& M^2_{eff}=\frac{g^2 \,T^2}{8} \label{HTLpart}
\end{eqnarray}
This HTL contribution is very similar to that of the gauge
fields. In fact including the gauge field is tantamount to replacing $g^2
\rightarrow g^2+e^2$ in the above expression.

In the HTL  approximation, the effective quark propagator has
poles describing two branches of collective
modes \cite{weldon,pis}. In this approximation the self
energy only has an absorptive part below the light cone, so that to leading
order in the HTL limit the on-shell width of the collective modes
vanishes. The effective mass for the collective modes $M^2_{eff}$ does not
break chiral symmetry \cite{weldon,pis,kapusta,lebellac,blaizot} but serves as
an infrared regulator. In order to obtain the width of the quasiparticles to
cure the pinch singularities we must go beyond the HTL
contributions. 


Although including the damping of excitations in the medium and consequently a
width in their spectral densities is a physically reasonable manner to cure the
pinch singularities, an alternative approach has been discussed in the
literature\cite{evans}.  


\subsection{Including the width - beyond HTL:}

There are several different contributions to the thermal width of the quark in
the medium. These come from scalars, pseudoscalars and gauge bosons and have
very different physical origins. We study separately the contributions from the
scalar and pseudoscalars and of gauge bosons to clarify the differences.

\subsubsection{Scalar/pseudoscalar contributions:} 

In the expression for the one loop self-energy given above
(\ref{epsilons})-(\ref{rhos}) the terms proportional to
$\delta(s-\omega_{k+q}+\bar{\omega}_{q})$ {\em could} lead to an imaginary part
on the mass shell of the collective modes. This would be the case if the delta
function is satisfied for $ s=\omega_{\pm}(k) $ with $ \omega_{\pm}(k)
$ being the
dispersion relation for the different branches of collective modes. This
relation clearly depends on the mass of the scalar or pseudoscalar particle in
the loop. Since this delta function arises from the process $ \phi
\leftrightarrow q+\bar{q} $ the constraint is satisfied if the
scalar (or pseudoscalar) boson {\bf can decay} into a $ q\bar{q} $ pair. 

The fact that the {\em decay} of the scalar leads to a thermal width for the
fermion has been previously recognized in\cite{shang} and can be understood
from a simple kinetic argument: the change in time of the spin-averaged quark
distribution can be written as a balance equation with a gain term minus a loss
term. The gain term arises at lowest order from the decay of the scalar into
$ q\bar{q} $ pairs with a factor 
$$ 
|{\cal M}_{fi}|^2 n_{k+q}(1-\bar{n}_k)(1-\bar{n}_q) , 
$$
while the loss term arises from the
`recombination' of quark-antiquark pairs into a sigma meson with probability
$$ 
|{\cal M}_{fi}|^2 (1+n_{k+q})\bar{n}_k \bar{n}_q 
$$ 
with $ |{\cal M}_{fi}|^2 $ the usual transition matrix element. The
linearization of the resulting Boltzmann 
equation leads to twice the damping rate\cite{shang} of the quark of momentum
$ k $. Thus the decay of the scalar into $ q\bar{q} $ pairs in the
medium induces a 
width for the quark or the collective excitations\cite{shang}.

In the strict chiral limit with $T<<T_c$, the mass of the $\sigma$ meson
$\approx 660\  \mbox{MeV}$ while that of the  constituent quark is $
\approx 300\  
\mbox{MeV}$ so that the kinematical decay condition  {\em could} be fulfilled. 
Near the critical temperature the sigma mass vanishes as does the
constituent quark mass (in the strict chiral limit) but now for $
T>>m_q $ the
important mass scale is determined by $ M_{eff} $ which for $ g
\approx 1.5 \; , 
\; T\approx T_c \approx 150\ \mbox{MeV} $ yields $ M_{eff}\approx 80\
\mbox{MeV} $. Whether the constraint could be fulfilled requires a detailed
study of the masses and their temperature corrections for the quarks and the
scalar mesons, a task beyond the scope of this article. 
However for $T\approx T_c$ where in the strict chiral limit $ m_q =0 $
and $ M_{eff} 
\approx 80-100\ \mbox{MeV} $, the sigma meson will be able to decay into
collective excitations giving the quarks a thermal width of origin very
different from the more familiar collisional width. From the one-loop
expression for the scalar and pseudoscalar contributions to the quark
self-energy we see that such a width will be of order $ g^2 \,T $ (for
a detailed 
expression for the width see\cite{shang}). In the perturbative analysis this
width is much smaller than the effective mass of the collective excitations
$ \approx g \,T $.

If the temperature dependent mass of the sigma prevents its decay
into collective modes by kinematics, a collisional contribution to the width of
the quark appears at two loops in the quark self-energy.  The argument given
above is valid whenever the fermion momentum in the self-energy loop
is $ >>g \,T $.
In particular, for $ T<T_c $ but for $ m_{\sigma}(T)>>g \,T $, the
on-shell delta 
function that determines the width requires that the momentum of the fermion in
the self-energy loop be $ > g \, T $ \cite{shang} and the use of the bare quark
propagator is justified. On the other hand very near the critical temperature
whenever $ m_{\sigma}(T)<<g \, T $ for soft external quark momentum
the internal 
quark line will also be soft in the kinematic region that contributes to the
width. In this case a full HTL resummed quark internal line must
be used. The analysis of such case has been provided in detail by
Thoma\cite{thoma} who concluded that the width of the quark in this limit is
also given by $\approx g^2 T$. This result {\em includes} the collisional width
but in the HTL approximation for the intermediate fermion in the
self-energy diagram \cite{thoma}.

In summary, for scalars (and pseudoscalars) the thermal width of the quark
collective modes has two different contributions both of order $ g^2
\,T $: the
decay of the scalar (sigma) into pairs of collective modes if kinematically
allowed and collisional damping in the HTL limit.

\subsubsection{Gauge boson contribution:}

The fermion damping rate arising from gauge boson exchange has been studied in
detail in the HTL approximation by several
authors \cite{pisarski,iancu,boyarg}. The gauge contribution to the quark
self-energy presents infrared divergences in bare perturbation theory. The
infrared divergences associated with the exchange of longitudinal gauge bosons
are a result of small angle Coulomb scattering and at zero temperature are
those of Rutherford scattering. At finite temperature the longitudinal gauge
boson (instantaneous Coulomb interaction) is screened by the Debye screening
mass $m_D \propto e\;T$ which cuts off the infrared and leads to a finite
contribution to the damping rate from longitudinal gauge bosons. For soft
external momentum this contribution has been found to be given
by\cite{pisarski,iancu}
\begin{equation}
\Gamma^{l,g}(k)= \alpha\;  A\;T~,\label{longi}
\end{equation}
with $A$ a constant that can be found numerically \cite{pisarski} and in QCD is 
of $ {\cal O}(1)$ and $\alpha = e^2/4\pi$. 

For a fermion excitation at {\em rest}  
the damping rate has been computed in \cite{pisarski,brapis,rebhan1} and
found to be given by 
\begin{equation}
\Gamma^{g}(k=0) = \alpha\; C\;T~,  \label{restgamma}
\end{equation}
with $C$ again being a numerical constant of ${\cal
O}(1)$\cite{pisarski,brapis}.

The detailed analysis of Refs.~\cite{pisarski,brapis,rebhan1} in QCD
lead to the result that the damping rate of moving quasiparticles 
with momenta $ k\gg  e^2 T $ in a non-abelian plasma are given by 
(up to an overall constant that depends on the gauge group structure)
\begin{equation}
\Gamma^g_{\pm}(k) = \frac{e^2 \,T}{4\pi}|v_{\pm}(k)|\;
\log\frac{1}{e}~,\label{gammaqcd} 
\end{equation}
where $v_{\pm}(k)$ are the group velocities for the two different branches of
collective modes. This expression is {\em not} valid for $k=0$ where the
damping rates of the fermion and plasmino at rest do not have the logarithmic
behavior in terms of the gauge coupling \cite{pisarski,brapis}, nor near
$k=k_{min}$ where the group velocity of the plasmino branch vanishes. Since the
HTL structure in QED is similar (up to gauge group factors) to
that of QCD (and scalar QED), Pisarski\cite{pisarski} suggested that a similar
form of the damping rate should be valid in a QED plasma, despite the fact that
there is no magnetic screening in the Abelian theory.

In QED the transverse photon propagator is only {\em dynamically} screened 
via Landau damping and the infrared divergences remain, possibly to all
orders in perturbation theory. More
recently\cite{iancu} a detailed study of the fermion propagator in the
Bloch-Nordsieck 
(eikonal) approximation in {\em real time} revealed that for $\omega_D\;t\;
|v_{\pm}(k)| \gg 1$   
\begin{equation}
S_k(t) \approx e^{-\alpha T |v_{\pm}(k)|\; t \ln(\omega_D\;t\;
|v_{\pm}(k)|)}~,  \label{nonexpo} 
\end{equation}
with $\omega_D \approx eT$ the Debye frequency and $v_{\pm}(k)$ the
group velocity  
of the fermionic collective modes. Although this
is not an exponential relaxation that would emerge from a Breit-Wigner
resonance shape of the spectral density, it does reveal
a particular time scale from which a damping rate  can be extracted
and is given by 
\begin{equation}
\Gamma^g_{\pm}(k) \approx \alpha\; T\; |v_{\pm}(k)|
\ln\frac{1}{e}~.\label{infrarate} 
\end{equation} 

This result has  been recently found in scalar QED (which has the
same HTL structure as QED and QCD to lowest order) within
 the dynamical renormalization group method\cite{boyarg}.  

At this stage it is important to describe the physics that leads to the damping
rate from the gauge boson contribution\cite{pisarski,iancu,boyarg}, which is
rather different from that of scalars found before.  The constraints from
energy momentum conservation in the imaginary part of the self-energy can only
be satisfied below the light cone where the HTL resummed gauge
boson propagator has support that arises from the Landau damping cut.  The
infrared process that leads to the non-exponential relaxation is the emission
and absorption of soft photons at almost right angles with the moving
fermion\cite{pisarski,iancu}.

Thus we summarize the contribution to the damping rate of the
collective excitations by the HTL resummed gauge boson exchange:
\begin{eqnarray}
&& \Gamma^g_{\pm}(k) \approx \alpha\; T\; |v_{\pm}(k)|\;
\ln\frac{1}{e}\quad \text{for} 
\quad k \gg g^2 \,T~,\label{movingrate} \\
&& \Gamma^g_{\pm}(k)\approx\alpha\; T\quad\quad\quad\quad\quad\quad~\;
\text{for}\quad k=0~. \label{restrate}
\end{eqnarray} 

The HTL resummed retarded and advanced propagators for the soft quarks can be
summarized as follows (here we also include the gauge contribution to the
HTL self-energy)
\begin{eqnarray}
\nonumber\\
S_{R}(k_0,\vec k)&=&
\frac{\gamma^0 \;A-\vec\gamma\cdot\hat k \; B}{A^2-B^2}\;,\\\nonumber\\
S_{A}(k_0,\vec k)&=&\frac{\gamma^0 \;A^*-\vec\gamma\cdot\hat k \; B^*}
{A^{*2}-B^{*2}}\quad ,\\\nonumber\\
A\pm B &\equiv&
k_0\pm k- i\epsilon-\frac{(g^2+e^2) \,T^2}{8k}\left[\mp 1\pm\frac{k_0\pm k}{2k}
\ln\frac{k_0+k- i\epsilon}{k_0-k-
i\epsilon}\right]+\Pi_{\pm}(k_0,\vec k)\;.
\\\nonumber
\end{eqnarray}
where $\Pi_{\pm}(k_0 ,\vec{k})$ are the ${\cal O}\left(g^2 T, \alpha T, \alpha
T \ln\frac{1}{e}\right)$ corrections to the HTL
self-energy that determine the width of the quark.  The retarded and advanced
propagators uniquely determine the time-ordered propagator via the spectral
representation:
\begin{eqnarray}
\nonumber\\
S^{++}=S_R+S_A+\left[1-2n_f(|k_0|)\right]\; {\mbox{sgn}}(k_0)\;(S_R-S_A)\quad.
\\\nonumber
\end{eqnarray}

The anomalous vertex with dressed fermion propagators is shown in
Fig. 3. 
Using the general expression (\ref{seffk}), incorporating the on-shell width
$ \sim g^2 \,T $ into our definition for $A$ and assuming that the external
momenta are negligible compared to all other scales in the problem we find
that the $ \pi \; A_\mu^+ \;A_\nu^+ $ vertex is given by:
\begin{eqnarray}
\nonumber\\
&&S^{+++}_{\pi\gamma\gamma}\approx 64m_q\;e^2g\int 
\frac{d^4p_1}{(2\pi)^4}\frac{d^4p_2}{(2\pi)^4}\times
\left\{i \;\epsilon^{\alpha\beta\mu\nu}\;p_{1\alpha} \;p_{2\beta}\;
\tilde\pi^{+}(-p_1-p_2) \;
\tilde A_\mu^+(p_1) \; \tilde{A}_\nu^+(p_2)\right.\label{htlvertex}
\\\nonumber\\\nonumber
&&\left.\int\frac{d^4k}{(2\pi)^4}
\left[{\mbox{Re}}\frac{1}{A^2(k_0,\vec k)-B^2(k_0,\vec k)}+
[1-2n_f(|k_0|)]{\mbox {sgn}}(k_0){\mbox{Im}}\frac{1}
{A^2(k_0,\vec k)-B^2(k_0,\vec k)}
\right]^3\right\}\\\nonumber
\end{eqnarray}

The infrared region for the real part is cut-off at the scale $(g^2+e^2)T$
where we have included the HTL contribution from the gauge fields, and the
imaginary part now has support below the light cone arising from the HTL
and also near the mass shell of the
collective modes because of the width $\Gamma \approx (g^2 \,T,\alpha
T, \alpha T \ln\frac{1}{e})$ which now cuts off the pinch singularity. The
real time vertex is now infrared finite and free of pinch singularities and
the local limit of zero frequency-momentum in the quark loop can be taken
safely.


At this point we must make an important comment -- in considering contributions
beyond the leading order in the HTL program, we must also include consistently
the vertex corrections. There are in fact two
sources of vertex corrections coming from $\sigma,\pi$ and photon exchange. As
long 
as $g>>e$, it is the former that is more important.
The inclusion of the photon exchange is necessary as a matter of principle 
to ensure that the Ward identities are satisfied and gauge invariance is
respected.  
Including the vertex
correction 
along with the quasiparticle width makes a detailed evaluation of the effective
vertex an extremely difficult task, certainly beyond the scope of this article.
The conclusions that we will obtain are therefore tentative pending inclusion 
of the vertex corrections. 


As mentioned in the introduction one of our principal goals is to understand
the effective interaction of the neutral pion with gauge fields to assess the
possibility of enhancement of neutral pion condensates in the presence of {\em
classical} electromagnetic fields produced in peripheral collisions.
To obtain the interaction of the pseudoscalar with classical background fields
we simply replace the fluctuating fields in Eq. (\ref{seffk}) with c-number
expectation values or classical sources. This yields the following coupling of
the pion to {\em classical fields} $ \tilde A_\mu $, 
\begin{eqnarray}
\nonumber\\
&&S_{\pi\gamma\gamma}=-2e^2g\int 
\frac{d^4p_1}{(2\pi)^4}\frac{d^4p_2}{(2\pi)^4}
\tilde\pi^{+}(-p_1-p_2) \; \tilde A_\mu(p_1) \;\tilde{A}_\nu(p_2)
\int\frac{d^4k}{(2\pi)^4}\times\\\nonumber\\\nonumber
&&\left\{\mbox{Tr}[S^{++}(k-p_1)\gamma_5S^{++}(k+p_2)\gamma^\mu
S^{++}(k)\gamma^\nu]\right.\\\nonumber 
&&\left.\hspace{2.7 in}
-\mbox{Tr}[S^{-+}(k-p_1)\gamma_5S^{++}(k+p_2)\gamma^\mu S^{+-}(k)\gamma^\nu]
\right.\\\nonumber
&&\left.- \mbox{Tr}[S^{++}(k-p_1)\gamma_5S^{+-}(k+p_2)\gamma^\mu
S^{-+}(k)\gamma^\nu]\right.\\\nonumber
&&\left.\hspace{2.7 in}
+\mbox{Tr}[S^{-+}(k-p_1)\gamma_5S^{+-}(k+p_2)\gamma^\mu S^{--}(k)\gamma^\nu]
\right\}
\end{eqnarray}

Thus the interaction of the pion with classical gauge fields is determined by a
combination of all the different terms in the CTP effective action. Rewriting
this expression in terms of the propagators in the Keldysh notation
[eqs.(\ref{keldysh1})-(\ref{keldysh3})], we obtain 
\begin{eqnarray}
\nonumber\\
&&S_{\pi\gamma\gamma}=-2e^2g\int 
\frac{d^4p_1}{(2\pi)^4}\frac{d^4p_2}{(2\pi)^4}
\tilde\pi^{+}(-p_1-p_2)\tilde
A_\mu(p_1)\tilde{A}_\nu(p_2)\int\frac{d^4k}{(2\pi)^4}\\\nonumber\\\nonumber 
&&\left\{\mbox{Tr}[S_R(k-p_1)\gamma_5S_R(k+p_2)\gamma^\mu S_R(k)\gamma^\nu]
+\mbox{Tr}[S_A(k-p_1)\gamma_5S_A(k+p_2)\gamma^\mu S_A(k)\gamma^\nu]
+\right.\\\nonumber\\\nonumber
&&\left.+
\mbox{Tr}[S_A(k-p_1)\gamma_5S_F(k+p_2)\gamma^\mu S_A(k)\gamma^\nu]
+\mbox{Tr}[S_A(k-p_1)\gamma_5S_R(k+p_2)\gamma^\mu S_F(k)\gamma^\nu]
+\right.\\\nonumber\\\nonumber
&&\left.+\mbox{Tr}[S_F(k-p_1)\gamma_5S_R(k+p_2)\gamma^\mu S_R(k)\gamma^\nu]
\right\}\\\nonumber
\end{eqnarray}

The local limit of the pion-photon vertex is obtained by taking the external
frequency and momentum to vanish in the quark loop. After the HTL
resummation and accounting for the width of the intermediate quarks, this limit
is unambiguous and finite. Thus we find

\begin{eqnarray}
&&S_{\pi\gamma\gamma}= \nonumber\\\nonumber
&&16 \, m_q\;e^2g\int 
\frac{d^4p_1}{(2\pi)^4}\frac{d^4p_2}{(2\pi)^4}\times
\left\{i\, \epsilon^{\alpha\beta\mu\nu}\; p_{1\alpha}p_{2\beta}\;
\tilde\pi^{+}(-p_1-p_2)
\tilde A_\mu^+(p_1) \tilde{A}_\nu^+(p_2)\int\frac{d^4k}{(2\pi)^4}\times
\right.
\\\nonumber\\\nonumber
&&\times\left.\left[\frac{1}{(A^2-B^2)^3}+\frac{1}{(A^{*2}-B^{*2})^3}+
[1-2n_f(|k_0|)] \; {\mbox {sgn}}(k_0) \;\left[
\frac{1}{(A^2-B^2)^3}-\frac{1}{(A^{*2}-B^{*2})^3}\right]\right]\right\}\nonumber\\
&& + \mbox{vertex corrections}
\label{finfor}
\end{eqnarray}

From this result the strength of the local pion-photon
vertex can be extracted, 
which is free of ambiguities and non-singular. It is of the form

\begin{equation}
g_{\pi \gamma \gamma} = \frac{m_q}{T^2} \;  e^2 \; g \; {\cal
F}\left[g, \alpha, \alpha \log\frac{1}{e}, \cdots \right] \label{finalvertex}
\label{nontrivial}
\end{equation}


The function ${\cal F}$ is finite and dimensionless and certainly quite
complicated by the logarithmic dependence on the coupling of the width and by
the 
vertex correction. It is beyond the scope of this article to obtain a closed
or approximate expression for this function. The main point of writing the
eqn. (\ref{nontrivial})
is to emphasize that after curing the pinch singularities by including the width (and vertex corrections)
 the vertex vanishes at $T=T_c$ in the strict chiral limit. This is in
agreement with the results in \cite{rob}. The necessity of including HTL
corrections was pointed out 
by \cite{gelis}. In the present context, the CTP effective action requires the
inclusion of a width as well, in order to regulate the pinch
singularities. This is one of the important results of the present work.

A crude, order of
magnitude estimate for the vertex may be found in the narrow-width
approximation, assuming that the width $\Gamma\sim O(g^2T)$. Then, the
dominant temperature dependent term in $g_{\pi\gamma\gamma}$ is $\sim
e^2gm_q\times(T^2/\Gamma^4)$ which 
is $O(e^2/(g^7T^2))$. (The $T^2/(\Gamma^4)$ factor may be easily understood by
going back to the discussion after Eq. (2.23).)


Before we embark on the study of nonequilibrium effects it is convenient to
summarize the  conclusions for the equilibrium case

\begin{itemize}
\item{The real time interaction vertices for the pseudoscalar and photons 
obtained by integrating out the quarks in the real time formulation display
infrared divergences in the local limit when the bare quark
propagators are used with $ m_q=0 $.
These divergences are cured by an HTL resummation of the quark lines {\em and}
by considering contributions beyond HTL that give an on-shell width to the
collective excitations. The resummed quark propagators now give rise to an
unambiguous and finite local interaction vertex.}

\item{The computation of the effective vertex in {\em imaginary time} in terms
of bare propagators \cite{rob,robtyt,baier} cannot be used to discuss
either the decay amplitude of the neutral pion or any dynamical aspects. In
that 
computation the external Matsubara frequencies were set to zero and
therefore the analytic continuation necessary to obtain the real time vertices
cannot be performed. This computation cannot reveal the subtle infrared and
pinch singularities that arise from finite temperature contributions to the
discontinuities of the triangle diagram that are responsible for ambiguities in
the limit of soft external momenta. A resummation of the quark propagator
including both HTL contributions {\em and } the corrections that lead
to the width of the fermionic collective modes is necessary to
regulate the infrared divergences in the soft limit.} 

\item{The infrared and pinch singularities of the real time vertex imply
that the local limit of the pseudoscalar-photon vertex is very sensitive to
the details of long-wavelength physics and is therefore {\em non-universal} in
the sense that it depends on the details of the underlying model. This is in
striking contrast with the situation in the vacuum where the vertex in the
local limit is completely determined by the triangle anomaly and the
Adler-Bardeen `theorem'\cite{adbardeen} guarantees that the vertex does not
receive radiative corrections. At finite temperature the situation is 
more complicated. As originally observed in \cite{robtyt}, while the
anomaly equation is 
independent of temperature and does not receive radiative corrections,
the vertex itself is not determined solely by the anomaly. Our studies
in equilibrium complement those of Gelis \cite{gelis} and lead to the
conclusion that the vertex is sensitive to the infrared and thus to
the details of the low energy scalar and pseudoscalar sectors.}

\item{After the HTL resummation and accounting for higher order corrections
that determine  
the fermion width the local limit of the vertex is given by
Eq. (\ref{finalvertex}) 
which  vanishes in the strict chiral limit  $ m_q =0 $ in the
symmetric phase, confirming the results in
\cite{rob,robtyt}. Hence, despite the fact that the imaginary time
computation of the vertex had missed the infrared sensitivity, it does reveal
the {\em correct} behavior for the pseudoscalar-photon vertex in the strict
chiral limit near the phase transition, when the constituent quark mass
vanishes. }

\end{itemize}


The full benefits of the real-time effective action will be appreciated in a
truly out of equilibrium 
situation. In the earlier sections we reproduced previous {\em equilibrium}
results within this approach and pointed out certain important subtleties
pertaining to the appearance of pinch singularities. We now turn our attention
to the true non-equilibrium situation which is the focus of our work.


\section{$\pi\gamma\gamma$ out of equilibrium}

We expect that nonequilibrium phenomena will be the rule rather than the
exception in heavy ion collisions. In particular, in terms of DCC formation, we
might expect that since electromagnetic effects break isospin, they might be
able to bias the neutral to charged pion ratios in a way that could be detected
in an event by event analysis. 

The scenarios of formation of DCC-type configurations envisage a strongly
`quenched' phase transition, in which the cooling of the plasma occurs via
hydrodynamic expansion on a time scale far shorter than the equilibration scale
of long-wavelength fluctuations\cite{moredcc,nosdcc,raja,photop}. 
Within the setting of the 
linear sigma model this situation is modeled by suddenly changing the form of
the potential for the scalars (and pseudoscalars) from one with an unbroken
symmetry minimum to another that displays symmetry breaking minima. 

While most of the previous studies focussed on an effective theory of scalars
and 
pseudoscalars (mainly the linear sigma model), in order to study the effects of
the anomalous vertex on possible enhancements of neutral pion condensates
(constituent) quarks must be added to the model. Following 
phenomenologically motivated descriptions \cite{rob,robtyt} we adopt the
$ U(2)\otimes U(2) $ gauged constituent quark model. This is, of
course, a simple extension of the model of the last section, with an $
SU(2)_L\times SU(2)_R $ global symmetry with two quark flavours and $
N_c=3 $ colors, coupled to electromagnetism:
\begin{eqnarray}
{\cal L}=&&i\bar{\psi}(\partial\!\!\!\slash-ieQA\!\!\!\slash)\psi
-2g\bar\psi(\sigma\tau^0+i\vec\pi\cdot\vec{\tau}\gamma_5)\psi
-\frac{1}{4}F_{\mu\nu}F^{\mu\nu}+{\cal L}_{\pi}.\label{sigmalag1}
\end{eqnarray}
Here $(\tau^0,\vec{\tau})$ are the isospin generators proportional to the
Pauli matrices, satisfying the normalization conditions
 Tr$(\tau^a\tau^b)=\delta^{ab}/2 $, $ \tau^0={\bf 1}/2 $ and 
\begin{eqnarray}
\psi=\left(\begin{array}{c}u\\\nonumber
d\end{array}\right)
\end{eqnarray}
where we have suppressed all color indices. The $ U(1)_{em} $ charge
is given by $ Q= \tau^0/3+\tau^3 $. $ {\cal L}_{\pi} $ determines the
dynamics of the pions and the sigma field and is a gauged $ O(4) $
scalar theory given by,  
\begin{eqnarray}
&&{\cal L}_{\pi}=
\frac{1}{2}(\partial_{\mu}\sigma)^2 + \frac{1}{2}
(\partial_{\mu}\vec{\pi})^2 + \frac{e^2}{2} (\pi_1^{2}+
\pi_2^{2})A_{\mu}A^{\mu}+j_{\mu}A^{\mu}-{\lambda}
(\sigma^2+\vec\pi^2-v_0)^2 +h\sigma
\label{veff}
\end{eqnarray}
\begin{equation}
j_{\mu}= e\; (\pi_1 \partial_{\mu}\pi_2-\pi_2\partial_{\mu}\pi_1)
\label{electrocurrent}
\end{equation}

Our approach to the problem will be to consider the evolution in time of the
chiral order parameter/condensate $\langle \sigma \rangle$, (where the
expectation value is 
taken in the time evolving density matrix), and to study how this evolution
feeds 
back into the fermion propagators in the triangle diagram. 

We consider the initial state to be described by free fields in
thermal equilibrium at a temperature $ T_i> T_c $ and
evolve the equations of motion with the $ T=0 $ masses. 
This `quench' scenario was originally described
in \cite{raja,boyadcc}. Several authors studied the cooling by hydrodynamic
expansion \cite{moredcc,nosdcc,photop} with similar results. Therefore
we consider preparing 
an initial density matrix that describes the system in equilibrium at an
initial temperature above the critical at an initial time $t_i$ which for
convenience will eventually be taken to be $t_i=0$. This initial density matrix
is now evolved in time with a potential corresponding to the zero temperature
situation\cite{boyadcc}.

At time $ t=t_i $ the temperature suddenly falls below the critical
value and the 
potential develops two (non-degenerate) minima.  The condensate now evolves in
time from the initial value towards the final equilibrium point with the
minimum (free) energy. We always begin with an equilibrium initial state where
the chiral order parameter has a non-zero expectation value $
\langle\sigma(x)\rangle\propto h $ which is the explicit symmetry
breaking term and yields a non-zero bare mass for the quarks. 

We should emphasize at the outset that the infrared and pinch singularities 
that plagued the equilibrium calculation are {\em absent} in this scenario. The
reasons for this are

\begin{enumerate}

\item We begin with a non-zero condensate value which yields a 
finite constituent quark mass at $ T_i>T_c $. This arises
from the explicit symmetry breaking in the linear sigma model, which is
responsible for the physical masses of the pions. 

\item The range of integration time 
is {\em finite}, from an initial time $t_i=0$ up to the time $t$ of
interest when the phase transition is complete, $ t \approx 10\
\mbox{fm}/c$. The fact that the 
time interval is finite introduces a natural infrared cutoff and prevents the
pinch singularities associated with the propagation of on-shell
intermediate states over very long times as emphasized in the non - equilibrium
situation \cite{bedaque}. That real time acts as an infrared cutoff for the propagation of
intermediate states has been understood within the framework of the damping of
collective excitations in a plasma in \cite{iancu,boyarg}. 
\end{enumerate}

Since the scalar couples to quarks and pions, the rolling of
$ \langle\sigma(x)\rangle = \sigma_0(t) $ has two important effects: i) an
effective {\em time dependent} mass for the pion fields that leads to
instabilities in the spinodal 
region of the potential and consequent growth of pion
fluctuations \cite{moredcc,nosdcc,raja,photop,boyadcc}, ii) a time
dependent constituent mass for the quarks.

These nonequilibrium effects result in the production of a large number of
pions  via spinodal growth of pion fluctuations
\cite{moredcc,nosdcc,raja,photop,boyadcc,largeN} and 
production of $q\bar{q}$ pairs through the time dependence of their
mass \cite{boyareheat}.  The quantum backreaction of quarks and pions affects
the dynamics of $ \langle\sigma(x)\rangle $ so that we cannot simply
focus on the 
classical equations of motion and a self-consistent treatment of the dynamics
of the condensate incorporating quantum effects is essential. Detailed
investigations of such dynamical situations in the $O(N)$ model and the scalar
sector of the linear sigma model using nonperturbative schemes such as the
large-$N$ limit have been carried out in  previous works by
various groups \cite{nosdcc,photop,boyadcc,largeN} without the
inclusion of quarks. 
The out of equilibrium evolution including fermion fields is treated
in \cite{boyareheat}.

As before, the full quantum field $ \sigma^{\pm}(x) $ can be split up into its
classical, c-number expectation value and quantum fluctuations about
such expectation value
\begin{eqnarray}
&&\sigma^{\pm}(\vec x, t)=\sigma_0(t)+\chi^\pm(\vec x, t)\\\nonumber
&&\langle\sigma^\pm(\vec x,
t)\rangle=\sigma_0(t) \quad , \quad \langle\chi^\pm\rangle=0 \; . 
\end{eqnarray}

The condensate will be introduced as a time-dependent mass term
for the quarks and its dynamics will be treated non-perturbatively. We split
the fermionic part of the Lagrangian into a free part and interaction
terms as follows: 
\begin{eqnarray}
&&{\cal L}_f={\cal L}_0+{\cal L}_{int}\nonumber\\\nonumber\\
&&{\cal L}_0=i\bar\psi^+[i\partial\!\!\!\slash-g\sigma_0(t)]\psi^+
-(+\longrightarrow-)\\\nonumber\\\nonumber
&&{\cal L}_{int}=e\bar\psi^+\gamma^\mu QA_\mu^+\psi^+
-ig\vec\tau\cdot\vec\pi^+\bar\psi^+\gamma_5\psi^+-(+\longrightarrow-)+\cdots.
\end{eqnarray}
where the dots stand for terms which are not relevant to the
pseudoscalar-photon vertex to one loop in the triangle diagram.  We
now combine the results of the (mean-field) large $N$ equations for
the linear sigma model \cite{moredcc,nosdcc,photop,boyadcc,largeN}
with those obtained for the Yukawa theory out of equilibrium
\cite{boyareheat} to obtain the self-consistent equations of motion in
the {\em mean field} approximation. 

\begin{eqnarray}
&& \ddot{\sigma}_0(t)-4\lambda\, v^2_0 \; \sigma_0(t)+4\lambda\,
\sigma^3_0(t)+4\lambda\, \sigma_0(t)\; \langle \vec{\pi}^2(x) \rangle -g\langle
\bar{\psi} \psi \rangle = h \nonumber\\\nonumber\\\nonumber
&& \langle \vec{\pi}^2(x) \rangle =
\frac{3}{2} \int \frac{d^3k}{(2\pi)^3}\;|\Phi_k(t)|^2
\coth\left[\frac{\omega_{k,\pi}(0)}{2T_i}\right] \\\nonumber\\\nonumber
&&\ddot{\Phi}_k(t)+\omega^2_{k,\pi}(t)\;\Phi_k(t) = 0 \; \; ; \; \;
\omega^2_{k,\pi}(t)= k^2 -4\lambda \; v^2_0 +4\lambda\;
\sigma^2_0(t)+4\lambda\; 
\langle \vec{\pi}^2(x) \rangle \nonumber \\\nonumber\\
&&[i\partial\!\!\!\slash-g\sigma_0(t)]\psi(x) = 0 \label{meanfieldeqns}
\end{eqnarray}
where we have set the number of pions $N=3$. The reader is referred
to\cite{moredcc,nosdcc,photop,boyadcc,boyareheat,largeN} for details
on these equations and their initial conditions.  

In the above equations we have neglected the contribution from the classical
gauge fields to the back-reaction problem for the dynamics of the condensate.
This is justified since the electromagnetic effects are of ${\cal O}(e^2)$ but
$\lambda, g >> e$ and $\lambda \approx g$
\cite{moredcc,nosdcc,photop,boyadcc,largeN}. This 
set of coupled self-consistent equations determine the time evolution of the
condensate including the large contributions from the pion and fermionic
fluctuations.  Although the integrals associated with $\langle \vec{\pi}^2
\rangle \; ; \;\langle \bar{\psi}\psi
\rangle $ are ultraviolet divergent and renormalization would be required, the
model under study is only useful as a low energy effective description with a
natural cutoff $\Lambda \leq 1\ \mbox{GeV}$.

The fermion propagators in the background of the condensate for a `quench'
scenario from an initial state with temperature $T_i$ are given
by\cite{boyareheat,kluger}:
\begin{eqnarray}
&& S_{\vec{k}}^{+ -}(t,t')=-
S_{\vec{k}}^{<}(t,t')=-i\int d^3x \; e^{-i\vec{k}\cdot
\vec{x}}\;  \langle \bar{\psi^-}(\vec{0},t') \psi^+(\vec{x},t)
\rangle \;,\label{ferplusmin}\\
&& S_{\vec{k}}^{- +}(t,t')=-
S_{\vec{k}}^{>}(t,t')=i\int d^3x \; e^{-i\vec{k}\cdot
\vec{x}} \; \langle \psi^-(\vec{x},t) \bar{\psi^+}(\vec{0},t')
\rangle \;,\label{ferminplus}\\\nonumber\\
&& S_{\vec{k}}^{++}(t,t')=S_{\vec{k}}^{>}(t,t')\Theta(t-t')
+S_{\vec{k}}^{<}(t,t')\Theta(t'-t) \;,\label{fertimeord}\\\nonumber\\
&& S_{\vec{k}}^{--}(t,t')=S_{\vec{k}}^{>}(t,t')\Theta(t'-t)
+S_{\vec{k}}^{<}(t,t')\Theta(t-t') \;,\label{ferantitimeord}\\\nonumber\\
&&S_{\vec{k}}^{>}(t,t')
=S_{0\vec k}^>(t,t^\prime)\left[1-n_{f}(\omega_k)\right]-
S_{0\vec k}^<(t,t^\prime)n_{f}(\omega_k)\;,\label{sgredef}
\\\nonumber\\
&&S_{\vec{k}}^{<}(t,t')= S_{0\vec k}^<(t,t^\prime)\left[1-n_f(\omega_k)\right]-
S_{0\vec k}^>(t,t^\prime)n_{f}(\omega_k)\;,
\label{slessdef}\\\nonumber\\
&&\omega_k=\sqrt{\vec{k}^2+g^2\sigma^2(0)}\;,\quad\quad\quad
n_{f}(\omega_k)=\frac{1}{e^{\beta \omega_k}+1}\;. \label{fermifactor}
\\\nonumber
\end{eqnarray} 

We have listed the nonequilibrium fermion propagators for the case where the
initial state is thermal with temperature $T_i=1/\beta$ and the chiral
condensate has the initial value $\sigma(0)$ for all times $t<t_i$. $S_{0\vec
k}^{>}$ and $S_{0\vec k}^<$ are the Wightman functions for the zero temperature
case and are given by
\begin{eqnarray}
&&S^>_{0k}(t,t')= \nonumber\\\nonumber
&&-i f_k(t)f^*_k(t')\left[{\cal{W}}_k(t)\gamma_0-
\vec{\gamma}\cdot
\vec{k}+g\sigma_0(t)\right]\left(\frac{1+\gamma_0}{2}\right)
\left[{\cal{W}}^*_k(t')\gamma_0-\vec{\gamma}\cdot
\vec{k}+g\sigma_0(t')\right]\;,\nonumber \\
&&S^<_{0k}(t,t')= \nonumber\\\nonumber
&&-i f^*_k(t)f_k(t')\left[{\cal{W}}^*_k(t)\gamma_0-
\vec{\gamma}\cdot
\vec{k}-g\sigma_0(t)\right]\left(\frac{1-\gamma_0}{2}\right)
\left[{\cal{W}}_k(t')\gamma_0-\vec{\gamma}\cdot
\vec{k}-g\sigma_0(t')\right]\;,\nonumber\\
&&\label{sgsl}\\
&&\quad {\cal{W}}_k(t)= i\; \frac{\dot{f}_k(t)}{f_k(t)}\;,
\label{timefreq1}\\
&&\left[\frac{d^2}{dt^2}+k^2+g^2\sigma_0^2(t)-ig\dot{\sigma}_0(t)\right]
f_k(t)=0\;, \label{fequation1}\\
&&f_k(t<t_i=0)= \frac{e^{-i\omega_k t}}{\sqrt{2\omega_k(\omega_k
+g\sigma_0(t_i=0) ) }} \;.\label{bcfoft1}
\end{eqnarray}
where the mode functions $f_k(t)$ determine the solution of the time dependent
Dirac equation. Details of the solution to the Dirac equation and the 
derivation of the nonequilibrium propagators for the fermions are provided in
Appendix B.  It is this dressing of the fermions by the time-dependent
condensate which will ultimately lead to nonequilibrium modifications of the
$\pi^0\gamma\gamma$ couplings. In terms of these propagators we
find\cite{boyareheat}
\begin{equation}
\langle \bar{\psi} \psi \rangle = \int \frac{d^3k}{(2\pi)^3}\;
\mbox{Tr}S_k^<(t;t)= 2 N_f N_c \int \frac{d^3k}{(2\pi)^3}
\left[1-2k^2|f_k(t)|^2\right] \label{psibarpsi}
\end{equation}
with $N_f,N_c$ the number of flavors (2) and of colors (3) respectively. As
mentioned in the introduction, it is  not  our goal here to fully
study  the coupled set of equations, but to focus on how this mean
field evolution provides nonequilibrium effects that modify the
pseudoscalar-photon coupling.

\subsection{Integrating out the fermions:}
As in the equilibrium case, the effective couplings will be induced once the
fermions are integrated out at the one-loop level via the triangle diagram. 
As before, we extract the nonequilibrium couplings by expanding
$\exp[iS_{int}]$ and 
integrating out the fermions at one-loop. This procedure automatically 
incorporates all the effects of the medium or the condensate which couples
to the fermions in the loops and we find the following expression for the
$\pi^0\gamma\gamma$ interactions in the nonequilibrium effective action:
\begin{eqnarray}
\nonumber\\
&&S_{\pi\gamma\gamma}^{noneq}=
-\frac{ge^2N_c}{3}\int
d^4x \; d^4x_1 \; d^4x_2\times\nonumber\\\nonumber\\\label{effint1}
&&\times\left\{\pi^{0+}(x)A_\mu^+(x_1)A_\nu^+(x_2)
\mbox{Tr}[S^{++}(x_2,x)\gamma_5 S^{++}(x,x_1)\gamma^\mu
S^{++}(x_1,x_2)\gamma^\nu] 
\right.\\\nonumber\\\nonumber 
&&\left.-\pi^{0+}(x)A_\mu^+(x_1)A_\nu^-(x_2)
\mbox{Tr}[S^{>}(x_2,x)\gamma_5S^{++}(x,x_1)\gamma^\mu S^{<}(x_1,x_2)\gamma^\nu]
\right.\\\nonumber\\\nonumber
&&\left.-\pi^{0+}(x)A_\mu^-(x_1)A_\nu^+(x_2)
\mbox{Tr}[S^{++}(x_2,x)\gamma_5S^{<}(x,x_2)\gamma^\nu S^{>}(x_2,x_1)\gamma^\mu]
\right.\\\nonumber\\\nonumber
&&\left.+\pi^{0+}(x)A_\mu^-(x_1)A_\nu^-(x_2)
\mbox{Tr}[S^{>}(x_2,x)\gamma_5S^{<}(x,x_2)\gamma^\nu S^{--}(x_2,x_1)\gamma^\mu]
\right\}.\\\nonumber
\end{eqnarray} 
where $S^>, S^<, S^{++}$ are the nonequilibrium fermionic propagators given by
(\ref{ferplusmin}-\ref{bcfoft1}) (see Appendix B for details).

\vspace{0.5in}
\centerline{\epsfbox{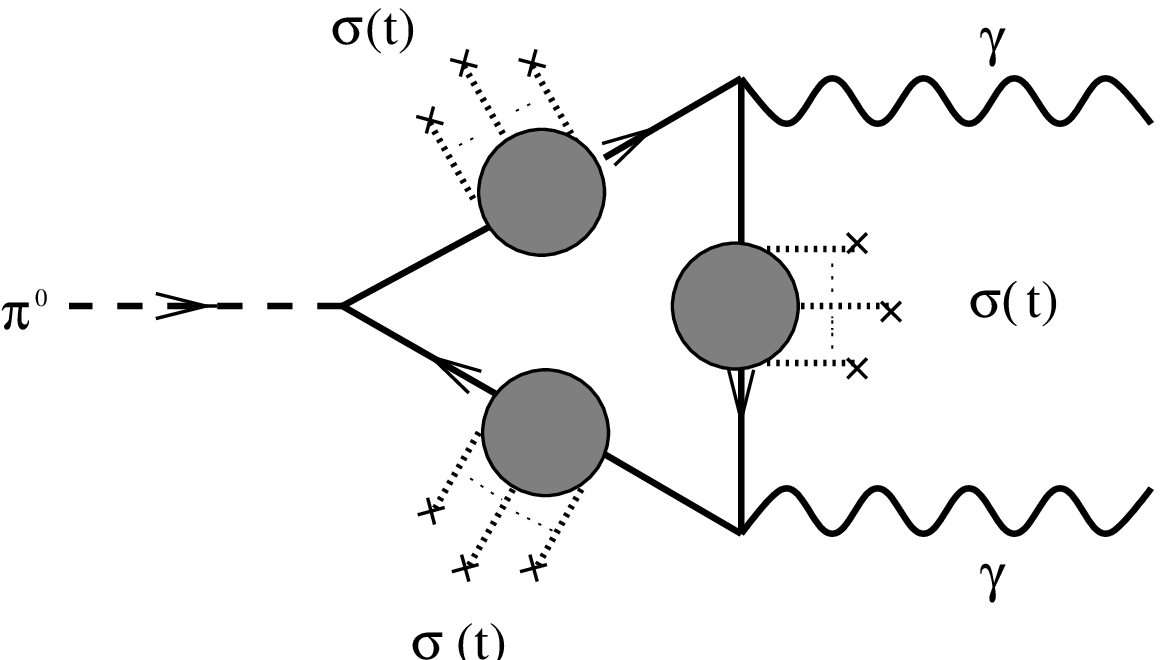}}
\vspace{0.5 in}

{FIG.4: {\small{The $\pi\gamma\gamma$ vertex out of equilibrium.}}}\\

Fig. 4, illustrates the nature of the anomalous couplings which are
induced out of equilibrium. Note that cross-couplings between fields
living on different branches of the CTP contour have been induced in
Eq. (\ref{effint1}). We note that if the zero temperature, equilibrium form for
the fermion propagators is used the cross-couplings vanish and only the
$\pi^{0+}A^{+}A^{+}$ vertex remains.  Now the effective interactions in
Eq. (\ref{effint1}) are manifestly non-local. In practice a local vertex is
obtained via a derivative expansion or an expansion in powers of the external
4-momenta where each derivative is suppressed by the mass of the fermions in
the loop which are being decoupled. In the usual equilibrium case the fermion
mass is simply $gv_0\approx gf_\pi$ and the derivative expansion is an
expansion in powers of $(\partial_\mu/gf_\pi)$ or an expansion in powers of
$(\partial_\mu/T)$ in the high temperature regime.

In the strongly out-of-equilibrium scenario under study, such a derivative
expansion is unwarranted. Furthermore, the lack of time translational
invariance due to the rolling of the condensate out of equilibrium prevents
Fourier transforms in time and a subsequent expansion in frequencies. The
Green's functions and consequently the kernels are no longer functions of time
differences, but in fact functions of the individual time arguments. It is also
clear that separating the center of mass and relative time variables as in the
case of a Wigner representation is of little use since there is no obvious
separation of time scales.

However if the electromagnetic field can be treated {\em classically} and if it
is slowly varying on the time scale of evolution of the condensate ($5-10\
\mbox{fm}/c$ \cite{moredcc,nosdcc,photop,boyadcc}) we can extract a
local interaction term of 
the type $\sim g_{\pi\gamma\gamma}(t) \; \pi^0 \; 
\vec E\cdot\vec B$ where the time-dependent coefficient determines the strength
of the anomalous coupling during the stage of nonequilibrium evolution of the
condensate. The restriction to static classical electromagnetic fields is
consistent with the assumptions used by \cite{asamuller} in their estimate of
anomalous enhancement of neutral pion condensates.

\subsection{Coupling to quasi-static $\vec E$ and $\vec B$ fields:}

Although the non-localities in the nonequilibrium vertex are difficult to
handle in general, couplings to quasi-static electromagnetic fields
are relatively easy to obtain and we will now outline the derivation
of these couplings. The assumption of quasi-static electromagnetic 
fields entails that their time variation is on scales much longer than those of
the nonequilibrium kernels ($\approx 5-10
\mbox{fm}/c$\cite{moredcc,nosdcc,photop,boyadcc})  that enter 
in the vertex. Under this assumption we can take these electromagnetic fields
outside of the time integral.   
We first introduce spatially Fourier-transformed fields:
\begin{eqnarray}
&&\tilde{\pi}^{0+}(\vec q,t)=\int d^3 x \;  e^{-i\vec q\cdot \vec x} \; 
\pi^{0+}(\vec x, t) \;  , \nonumber \\
&&\tilde{A}^{\pm}_{\mu}(\vec q,t)=\int d^3 x \;  e^{-i\vec q\cdot \vec x} \; 
A^{\pm}_{\mu}(\vec x, t) \; .
\end{eqnarray}
In terms of these fields, we find that the following anomalous interactions are
induced at one-loop in the nonequilibrium action:
\begin{eqnarray}
\nonumber\\
&&S^{noneq}_{\pi\gamma\gamma}=-e^2g\frac{N_c}{3}\int dt dt_1 dt_2 \int 
\frac{d^3q_1}{(2\pi)^3}\frac{d^3q_2}{(2\pi)^3}\times\label{effaction}\\\nonumber
&&\times\left[\tilde\pi^{0+}(\vec q_1,t)\tilde
A_\mu^+(\vec q_2,t_1) 
\tilde{A}_\nu^+(-\vec q_1-\vec q_2;t_2)
\int\frac{d^3k}{(2\pi)^3}V_{+++}^{\mu\nu}(\vec q_1+\vec q_2+\vec k,
\vec k+\vec q_2, \vec k;t,t_1,t_2) 
+\right.\\\nonumber\\\nonumber
&&\left.-\tilde\pi^{0+}(\vec q_1,t)\tilde
A_\mu^+(\vec q_2,t_1)\tilde{A}_\nu^-(-\vec q_1-\vec q_2;t_2)
\int\frac{d^3k}{(2\pi)^3}V_{++-}^{\mu\nu}(\vec q_1+\vec q_2+\vec k, 
\vec k+\vec q_2, \vec k;t,t_1,t_2)+\right.\\\nonumber\\\nonumber
&&\left.-\tilde\pi^{0+}(\vec q_1,t)\tilde
A_\mu^-(\vec q_2,t_1)\tilde{A}_\nu^+(-\vec q_1-\vec q_2;t_2)
\int\frac{d^3k}{(2\pi)^3}V_{+-+}^{\mu\nu}(\vec q_1+\vec q_2+\vec k, 
\vec k+\vec q_2, \vec k;t,t_1,t_2)+\right.\\\nonumber\\\nonumber
&&\left.+\tilde\pi^{0+}(\vec q_1,t)\tilde
A_\mu^-(\vec q_2,t_1)\tilde{A}_\nu^-(-\vec q_1-\vec q_2;t_2)
\int\frac{d^3k}{(2\pi)^3}V_{+--}^{\mu\nu}(\vec q_1+\vec q_2+\vec k, 
\vec k+\vec q_2, \vec k;t,t_1,t_2)\right]\\\nonumber
\end{eqnarray}
where, $V_{+++}^{\mu\nu}$, $V_{++-}^{\mu\nu}$, $V_{+-+}^{\mu\nu}$, and 
$V_{+--}^{\mu\nu}$ are given by the following expressions in terms of the
nonequilibrium propagators,
\begin{eqnarray}
\nonumber\\
&&V^{\mu\nu}_{+++}(\vec k_1,\vec k_2,\vec k)=\label{v+++}\\\nonumber\\\nonumber
&&\mbox{Tr}[S^{>}_{k_1}(t_2,t)\gamma_5S^{>}_{k_2}(t,t_1)\gamma^\mu 
S^{<}_{k}(t_1,t_2)\gamma^\nu]\Theta(t_2-t)\Theta(t-t_1)\Theta(t_2-t_1)+
\\\nonumber\\\nonumber
&&\mbox{Tr}[S^{<}_{k_1}(t_2,t)\gamma_5S^{<}_{k_2}(t,t_1)\gamma^\mu 
S^{>}_{k}(t_1,t_2)\gamma^\nu]\Theta(t-t_2)\Theta(t_1-t)\Theta(t_1-t_2)+
\\\nonumber\\\nonumber
&&\mbox{Tr}[S^{>}_{k_1}(t_2,t)\gamma_5S^{<}_{k_2}(t,t_1)\gamma^\mu 
S^{>}_{k}(t_1,t_2)\gamma^\nu]\Theta(t_2-t)\Theta(t_1-t)\Theta(t_1-t_2)+
\\\nonumber\\\nonumber
&&\mbox{Tr}[S^{>}_{k_1}(t_2,t)\gamma_5S^{<}_{k_2}(t,t_1)\gamma^\mu 
S^{<}_{k}(t_1,t_2)\gamma^\nu]\Theta(t_2-t)\Theta(t_1-t)\Theta(t_2-t_1)+
\\\nonumber\\\nonumber
&&\mbox{Tr}[S^{<}_{k_1}(t_2,t)\gamma_5S^{>}_{k_2}(t,t_1)\gamma^\mu 
S^{>}_{k}(t_1,t_2)\gamma^\nu]\Theta(t-t_2)\Theta(t-t_1)\Theta(t_1-t_2)+
\\\nonumber\\\nonumber
&&\mbox{Tr}[S^{<}_{k_1}(t_2,t)\gamma_5S^{>}_{k_2}(t,t_1)\gamma^\mu 
S^{<}_{k}(t_1,t_2)\gamma^\nu]\Theta(t-t_2)\Theta(t-t_1)\Theta(t_2-t_1)\;,
\\\nonumber\\\nonumber\\
&&V^{\mu\nu}_{++-}(\vec k_1,\vec k_2,\vec k)=\label{v++-}\\\nonumber
\\\nonumber
&&\mbox{Tr}[S^{>}_{k_1}(t_2,t)\gamma_5S^{>}_{k_2}(t,t_1)\gamma^\mu 
S^{<}_{k}(t_1,t_2)\gamma^\nu]\Theta(t-t_1)+
\\\nonumber\\\nonumber
&&\mbox{Tr}[S^{>}_{k_1}(t_2,t)\gamma_5S^{<}_{k_2}(t,t_1)\gamma^\mu 
S^{<}_{k}(t_1,t_2)\gamma^\nu]\Theta(t_1-t)\;,\\\nonumber\\\nonumber\\
&&V^{\mu\nu}_{+-+}(\vec k_1,\vec k_2,\vec k)=\label{v+-+}\\\nonumber\\\nonumber
&&\mbox{Tr}[S^{>}_{k_1}(t_2,t)\gamma_5S^{<}_{k_2}(t,t_1)\gamma^\mu 
S^{>}_{k}(t_1,t_2)\gamma^\nu]\Theta(t_2-t)+
\\\nonumber\\\nonumber
&&\mbox{Tr}[S^{<}_{k_1}(t_2,t)\gamma_5S^{<}_{k_2}(t,t_1)\gamma^\mu 
S^{>}_{k}(t_1,t_2)\gamma^\nu]\Theta(t-t_2)\;,\\\nonumber\\\nonumber\\
&&V^{\mu\nu}_{+--}(\vec k_1,\vec k_2,\vec k)=\label{v+--}\\\nonumber\\\nonumber
&&\mbox{Tr}[S^{>}_{k_1}(t_2,t)\gamma_5 S^{<}_{k_2}(t,t_1)\gamma^\mu 
S^{>}_{k}(t_1,t_2)\gamma^\nu]\Theta(t_2-t_1)+
\\\nonumber\\\nonumber
&&\mbox{Tr}[S^{>}_{k_1}(t_2,t)\gamma_5 S^{<}_{k_2}(t,t_1)\gamma^\mu 
S^{<}_{k}(t_1,t_2)\gamma^\nu]\Theta(t_1-t_2)\;,
\\\nonumber\\\nonumber
&&{\mbox {and}}\;\;\vec k_1=\vec k+\vec q_1+\vec q_2,\; \vec k _2=\vec k+\vec
q_2. 
\end{eqnarray}

We now assume that the photons can be treated as classical quasi-static
background fields  in accord with our original goal of studying the
response of the 
neutral pion field to large electromagnetic fields in peripheral collisions and
in agreement with the treatment of \cite{asamuller,photop}. We then
perform a spatial derivative expansion to extract a local
coupling. The result will be 
gauge-invariant under restricted (static) gauge transformations.  While such
couplings cannot be used to obtain the pion {\em width} these are relevant for
addressing the possibility of neutral pion condensates induced by the
electromagnetic fields produced in peripheral collisions or through the phase
transition.

We introduce c-number expectation values (mean fields) for the
electromagnetic fields, 
\begin{eqnarray}
A_\mu^{classical}=\langle A_\mu^{\pm}\rangle={\cal A}_{\mu}.
\end{eqnarray}
In the presence of these classical sources the coupling of $\pi^0$ to the
electromagnetic fields takes the following form,
\begin{eqnarray}
\nonumber\\
&&-e^2g\frac{N_c}{3}\int_{t_i}^\infty dt \int
\frac{d^3 q_1}{(2\pi)^3}\frac{d^3q_2}{(
2\pi)^3}\;\;\tilde\pi^{0+}
(\vec q_1,t)\;\tilde {\cal A}_\mu(\vec q_2)\;\tilde {\cal A}_\mu
(-\vec q_1-\vec q_2)\times\label{voft}\\\nonumber
&&\times\int_{t_i}^\infty dt_1\int_{t_i}^\infty dt_2\int \frac{d^3k}{(2\pi)^3}
(V^{\mu\nu}_{+++}-V^{\mu\nu}_{++-}-V^{\mu\nu}_{+-+}
+V^{\mu\nu}_{+--}).
\\\nonumber
\end{eqnarray}

This is obtained from Eq. (\ref{effaction}) by replacing the fluctuating
gauge fields 
with their c-number expectation values. The range of the time
integrals in the expressions above is from  $ t_i=0 $ to $ t $ as
determined by the step-functions (see below). 

In this nonequilibrium scenario, the anomalous coupling to classical
fields is then
\begin{eqnarray}
\nonumber\\
e^2g\frac{N_c}{3}\int dt \int
\frac{d^3 q_1}{(2\pi)^3}\frac{d^3q_2}{(
2\pi)^3}\;\;\tilde\pi^{0+}
(\vec q_1,t)\;\tilde {\cal A}_\mu(\vec q_2)\;\tilde {\cal A}_\mu
(-\vec q_1-\vec q_2)\times V^{\mu\nu}(\vec q_1,\vec q_2;t).
\end{eqnarray}
where 
\begin{eqnarray}
&&V^{\mu\nu}(\vec q_1,\vec q_2;t)=\int_{t_i}^{\infty}dt_1
\int_{t_i}^{\infty} dt_2 \int
\frac{d^3k}{(2\pi)^3}\times\nonumber\\\nonumber\\\nonumber 
&&\left[-\left\{\mbox{Tr}[S^{>}_{k_1}(t_2,t)
\gamma_5S^{>}_{k_2}(t,t_1)\gamma^\mu  
S^{<}_{k}(t_1,t_2)\gamma^\nu]+
\mbox{Tr}[S^{<}_{k_1}(t_2,t)\gamma_5S^{<}_{k_2}(t,t_1)\gamma^\mu 
S^{>}_{k}(t_1,t_2)\gamma^\nu]\right\}\right.\\\nonumber\\\nonumber
&&\left.\times\Theta(t-t_1)\Theta(t-t_2)+\right.\\\nonumber\\\nonumber
&&\left.+\left\{\mbox{Tr}[S^{>}_{k_1}(t_2,t)\gamma_5S^{<}_{k_2}(t,t_1)
\gamma^\mu  
S^{>}_{k}(t_1,t_2)\gamma^\nu]
+\mbox{Tr}[S^{>}_{k_1}(t_2,t)\gamma_5S^{<}_{k_2}(t,t_1)\gamma^\mu 
S^{<}_{k}(t_1,t_2)\gamma^\nu]\right\}\times\right.\\\nonumber\\\nonumber
&&\left.\Theta(t-t_1)\Theta(t-t_2)
\Theta(t_2-t_1)+\right.\\\nonumber\\\nonumber
&&\left.+\left\{\mbox{Tr}[S^{<}_{k_1}(t_2,t)\gamma_5S^{>}_{k_2}(t,t_1)
\gamma^\mu  
S^{>}_{k}(t_1,t_2)\gamma^\nu]
+\mbox{Tr}[S^{<}_{k_1}(t_2,t)\gamma_5S^{>}_{k_2}(t,t_1)\gamma^\mu 
S^{<}_{k}(t_1,t_2)\gamma^\nu]\right\}\times\right.\\\nonumber\\\nonumber
&&\left.\times\Theta(t-t_2)\Theta(t-t_1)\Theta(t_2-t_1)\right]\;,
\end{eqnarray}
where, as noted before $\vec k_1=\vec k+\vec q_1+\vec q_2$ and $\vec k_2 = \vec
k+\vec q_2$.

We now expand the above traces in 
powers of $\vec q_1$ and $\vec q_2$ which corresponds to performing a {\em
spatial} derivative expansion. Furthermore since we are confining ourselves
to interactions with static gauge fields it suffices to retain only those
terms in the traces which yield couplings of the type
$\;\pi^0\epsilon^{ijk}\;\partial_i{\cal A}_0\;\partial_j{\cal A}_k$. After
considerable but straightforward algebra, the details of which are not
particularly illuminating we 
find that the local effective vertex for the neutral pion coupling to static,
classical $\vec E$ and $\vec B$ fields is given by 
\begin{eqnarray}
\nonumber\\
S^{noneq}_{\pi\gamma\gamma}
=-4\int d^4 x \;\;
g_{\pi\gamma\gamma}(t)\;\epsilon^{ijk}\;\pi^{0+}(\vec x,t)
\;\partial_i {\cal A}_0(\vec x)\;\partial_j{\cal A}_k(\vec x)
\label{effevertex}\\\nonumber
\end{eqnarray}
where the amplitude or strength of the time-dependent anomalous
coupling is:
\begin{eqnarray}
\nonumber\\
&&g_{\pi\gamma\gamma}(t)=-\frac{g e^2 N_c}{6}
\int_{t_i}^{\infty}dt_1\int_{t_i}^{\infty}
dt_2\int \frac{d^3k}{(2\pi)^3}\times\label{goft}\\\nonumber\\\nonumber
&&\times\left[\left\{X^{-*}_k(t_1,t_2)\;\;
\left(1+\frac{2k^2}{3}\frac{\partial}{\partial k^2}\right)\;\;
\left[Y_k(t_2,t)\;Y_k(t,t_1)+Z_k(t_2,t)\;Z_k(t,t_1)\right]+\right.\right.
\\\nonumber 
\\\nonumber
&&\left.\left.-\frac{2k^2}{3}\;\;Y_k^*(t_1,t_2)\;\;
\left[Y_k(t,t_1)\;\frac{\partial}{\partial k^2}X^-_k(t_2,t)
+Y_k^*(t,t_2)\;\frac{\partial}{\partial k^2}X^{-*}_k(t_1,t)+
\right.\right.\right.\\\nonumber\\\nonumber
&&\left.\left.\left.-Z_k(t,t_1)\;\frac{\partial}{\partial k^2}
X^+_k(t_2,t)-Z_k^*(t,t_2)\;\frac{\partial}{\partial
k^2}X^{+*}_k(t_1,t)\right]\right\}\;
\Theta(t-t_1)\Theta(t-t_2)+\right.\\\nonumber\\\nonumber
\\\nonumber
&&\left.+\left\{X^{-}_k(t_1,t_2)\;\;
\left(1+\frac{2k^2}{3}\frac{\partial}{\partial k^2}\right)
\;\;
\left[-Y_k(t_2,t)\;Y_k^*(t,t_1)+Z_k(t_2,t)\;Z_k^*(t,t_1)\right]+\right.\right.
\\\nonumber\\\nonumber
&&\left.\left.+\frac{2k^2}{3}\;\;Y_k(t_1,t_2)\;\;\left[
Y_k^*(t,t_1)\;\frac{\partial}{\partial k^2} X^-_k(t_2,t)+Y_k^*(t,t_2)\;
\frac{\partial}{\partial k^2}X^{-}_k(t_1,t)+
\right.\right.\right.\\\nonumber\\\nonumber
&&\left.\left.\left.+Z_k^*(t,t_1)\; 
\frac{\partial}{\partial k^2}
X^+_k(t_2,t)+Z_k^*(t,t_2)\;
\frac{\partial}{\partial
k^2}X^{+}_k(t_1,t)\right]+
{\mbox{c.c.}}\right\}\times\right.\\\nonumber
\\\nonumber
&&\left.\Theta(t-t_1)\Theta(t-t_2)\Theta(t_2-t_1)\right].\label{noneqvertex}
\end{eqnarray}

The variables $ X_k^\pm,\;Y_k,\;Z_k $ are related to the mode
functions $ f_k $ via the following definitions:
\begin{eqnarray}
\nonumber\\
&&X_k^{-}(t,t^\prime)=f_k(t)f_k^*(t^\prime)
\left[S_k(t)S_k^*(t^\prime)-k^2\right](1-n_f)
-f_k^*(t)f_k(t^\prime)
\left[S_k^*(t)S_k(t^\prime)-k^2\right]n_f\label{xkminus}
,\\\nonumber\\
&&X^+_k(t,t^\prime)=f_k(t)f_k^*(t^\prime)
\left[S_k(t)S_k^*(t^\prime)+k^2\right]
(1-n_f)
+f_k^*(t)f_k(t^\prime)\left[S_k^*(t)S_k(t^\prime)+
k^2\right]n_f\label{xkplus}
,\\\nonumber\\
&&Y_k(t,t^\prime)=-f_k(t)f_k^*(t^\prime)
\left[S_k(t)+S_k^*(t^\prime)\right](1-n_f)-
f_k^*(t)f_k(t^\prime) \left[S_k^*(t)+S_k(t^\prime)\right]
n_f\label{yk}
,\\\nonumber\\
&&Z_k(t,t^\prime)=f_k(t)f_k^*(t^\prime)
\left[S_k^*(t^\prime)-S_k(t)\right]
(1-n_f)
+f_k^*(t)f_k(t^\prime)\left[S_k(t^\prime)-S_k^*(t)\right]n_f\label{zk}
,\\\nonumber\\
&&S_k(t)=i\; \frac{\dot{f}_k(t)}{f_k(t)}+g \; \sigma_0(t).
\end{eqnarray}
\noindent 

While these expressions look rather unwieldy, it is clear that the
$ \pi^0\gamma\gamma $ vertex receives important nonequilibrium corrections
through the evolution of the condensate. In the self-consistent mean-field
treatment advocated here, we see that the effective vertex out of equilibrium
receives contributions from {\em all the field fluctuations} through the
back-reaction. 

These are {\bf memory} effects since the anomalous coupling at time $
t $ depends on the fields at all times $ t' \leq t $.
Analogous effects are known to appear in related phenomena as in the
nonequilibrium dynamics of the condensate
\cite{moredcc,nosdcc,photop,boyadcc,largeN,houches} 
where pion spinodal fluctuations and the nonlinear backreaction
are very important. Hence, just as in the previous study 
of the effective vertex in equilibrium we conclude that, unlike the anomaly
equation (see Appendix A), the effective pseudoscalar-photon vertex 
receives important corrections from long-wavelength fluctuations and 
is not determined by the anomaly equation.

We are now in position to obtain the effective equations of motion for the
neutral pion field including the self-consistent mean field evolution as well
as the effective low energy interaction vertex with the classical
electromagnetic fields:
\begin{eqnarray}
&&\partial_{\mu}\partial^{\mu}\pi^0(\vec x,t)+M^2(t)\; \pi^0(\vec x,t)
= -g_{\pi \gamma \gamma}(t) \; 
\vec{E}(\vec x,t)\cdot \vec{B}(\vec x,t) \label{pi0eqn}\\\nonumber\\
&&M^2(t) =-4\lambda\; v^2_0 +4\lambda\; \sigma^2_0(t)+4\lambda \;\langle
\vec{\pi}^2(x) \rangle \label{massoft}  
\end{eqnarray}
where the last term in (\ref{massoft}) can be read from
Eq. (\ref{meanfieldeqns}).

This equation displays several important physical processes: i) early time
spinodal instabilities resulting from the negative contribution to the mass
term, just as in the more conventional treatments of formation of
DCC \cite{moredcc,nosdcc,photop,boyadcc,largeN}, ii) an inhomogeneity
induced by the classical 
quasi-static electromagnetic fields as a result of the anomalous coupling but
with a vertex that is a strong and rapid function of time. This vertex must be
obtained self-consistently from the coupled set of mean-field equations
(\ref{meanfieldeqns})-(\ref{fequation1}) by inserting the mode functions in
Eq. (\ref{noneqvertex}). Although we do not intend to offer a numerical
analysis of the resulting equations in this article, the nonequilibrium
equation of motion (\ref{pi0eqn}) has the potential for providing an
enhancement 
of electromagnetic isospin breaking effects through the combination of spinodal
instabilities and nonequilibrium time dependence of the pseudoscalar
vertex. This enhancement {\em could} lead to an experimentally observable
signal in the neutral to charged pion ratio. We postpone to a forthcoming
article the full numerical study of these equations and an assessment of the
potential phenomenological impact of the nonequilibrium dynamics.

\section{discussions, conclusions and outlook}

The axial anomaly displays a wealth of fascinating phenomena at finite
temperature as well as in out of equilibrium situations, perhaps even more
than at zero temperature. What we have done in this work is to use the
techniques of real time quantum field theory to understand the relationship
between the axial anomaly and the $\pi^0 \rightarrow \gamma \gamma$ amplitude
and develop a a systematic description of the in-medium
equilibrium and nonequilibrium corrections to the pseudoscalar-photon vertex.

The real time equilibrium calculation allowed us to make contact with some
previous results. In particular, we obtained agreement with the results in
\cite{gelis} by finding that the usual perturbative calculation of the 
triangle diagram in real time and finite temperature in terms of bare quark
propagators is afflicted by infrared divergences in the strict chiral limit of 
vanishing quark masses, and pinch singularities in the local limit of the
vertex. The latter singularities for vanishing external four momentum are a
manifestation of the propagation of undamped on-shell intermediate states in
the medium. 
Both types of divergences arise from the
soft momentum region for the internal quark lines. In this region the quark
propagators must be resummed including the HTL contribution from
scalars and gauge bosons. The pinch singularities, however, are not cured
by the HTL resummation but require going beyond the HTL limit and
accounting for the on-shell width of the collective excitations. 
These
conclusions are still somewhat tentative, since we have not included the
vertex corrections whose contributions can be significant (see for
e.g.\cite{carrington}). 

After the infrared and pinch singularities had been smoothed out by HTL
resummation and the inclusion of the quasiparticle width, we found that in the
strict chiral limit, the anomalous vertex vanishes when $m_q \rightarrow 0$ in
agreement with the original results of Pisarski \cite{rob,robtyt,baier}
and the suggestion by Gelis\cite{gelis}.

We next focused on the description of nonequilibrium dynamics during a
quenched chiral phase transition described by a constituent quark model with a
scalar-pseudoscalar sector described by the linear sigma model. The dynamics of
a quenched chiral phase transition is studied in a self-consistent mean-field
theory as advocated in many treatments of DCC formation
\cite{moredcc,nosdcc,raja,photop,boyadcc} but including the dynamics of the 
quarks. The nonequilibrium evolution of the chiral condensate introduces a
time dependent mass for the quarks. We obtained an analytic expression for the
quark propagators in the mean field limit and used these to compute
the triangle diagram out of equilibrium. The effective local
pseudoscalar-photon vertex out of equilibrium is then extracted for
quasistatic, classical electromagnetic field configurations as argued
to be relevant in peripheral ultrarelativistic heavy 
ion collisions. An interesting aspect of the nonequilibrium calculation is
that its treatment as an initial value problem has eliminated the infrared and
pinch singularities found in the equilibrium situation. 

From the nonequilibrium vertex we constructed the nonequilibrium equations of
motion 
for the neutral pion in the mean field approximation, including the
effective pseudoscalar-photon vertex.
This equation of motion revealed
potential enhancements of electromagnetic isospin breaking effects through
spinodal instabilities and the nonequilibrium anomalous vertex.
Thus the triangle diagram and the effective
non-equilibrium pseudoscalar-photon vertex is {\em sensitive} to
long-wavelength fluctuations of scalars and pseudoscalars through back-reaction
effects.

One of the important observations (originally made in \cite{robtyt}) of this
article is that whereas the anomaly 
equation 
and the axial Ward identity are independent of the model and the low energy 
sector of the theory, the pseudoscalar-photon vertex in the medium in or
out of equilibrium is {\em non-universal}, model dependent and very sensitive
to low energy physics.  
The next calculation to do would be a numerical study of the effective equation
of motion for the neutral pion, with an eye towards answering the question
addressed in \cite{asamuller}: can the anomalous vertex enhance the formation
of neutral pion DCC regions. This would have fascinating phenomenological
consequences. 

\section{acknowledgements:}
S. P. K. would like to thank L. Yaffe for several enlightening discussions and
comments at various stages of this work. He would also like to thank S. Jeon,
B. Muller, K. Rajagopal and M. Tytgat for helpful discussions. S. P. K. was
supported in part by the DOE grant DE-FG03-96ER40956. 
H. J. de V. thanks R. Pisarski and M. Tytgat for discussions.  
D.B.
thanks the NSF for partial support through grants PHY-9605186 and
INT-9815064 and LPTHE at Universit\'e Paris VI and VII where
part of this research was done for hospitality and support. D.B. thanks
R. Pisarski for an illuminating conversation a long time ago and in a galaxy
far away.  D.B., H. J. de V. and R. H. acknowledge support from NATO. R.H. was
supported in part by DOE grant DE-FG02-91-ER40682. 
\newpage
\appendix

\section{The Anomalous Ward identity out of equilibrium:}

The fact that the axial Ward identity is unmodified by thermal corrections in
a heat bath at finite temperature was explicitly shown in \cite{itomuller}. 
In this appendix we explicitly show via perturbation theory at 1-loop that the
axial Ward identity and the anomaly equation are not modified in or out of
equilibrium. This is a non-trivial statement in the real time formulation
since there are, as we will see below, different contributions to the
triangle diagram and anomaly equation arising from the different branches in
the CTP contour of the generating functional. 

 We will use the point-splitting approach
to derive the anomalous conservation law but implemented in real time.

The classically conserved axial current in the LSM is 
\begin{eqnarray}
&&j_{5\mu}^{a}=\bar\psi\gamma_\mu\gamma_5\tau^a\psi+
\sigma\partial_\mu\pi^a-\pi^a\partial_\mu\sigma \; \; ,
\end{eqnarray}
where the superscript $(a)$ denotes the isospin component under consideration.
The quantity of interest is the $a=3$ component of the current which exhibits
the anomalous behaviour. In the nonequilibrium CTP formulation all the
operators reside on the CTP contour, carrying $+$ and $-$ labels depending on
which part of the time contour they belong to and we will 
compute within perturbation theory the expectation value of 
\begin{eqnarray}
\langle\partial^\mu j^{3+}_{5\mu}\rangle.
\end{eqnarray}
which is defined on the forward contour.
In the subsequent steps we will omit the isospin index since it is  clear
which 
component we are focussing on. The expectation value above is of course {\it a
priori} ill-defined because of the usual short-distance singularities in the
operator product expansion. We therefore regulate this composite
operator by point-splitting in a gauge invariant manner as follows,
\begin{eqnarray}
\langle\partial^\mu j_{5\mu}^{(+)}(x)\rangle
\equiv \lim_{\epsilon^\mu\rightarrow 0}
\langle\partial^\mu \tilde j_{5\mu}^{(+)}(x,\epsilon)\rangle\;,
\end{eqnarray} 
where we have defined a new operator 
\begin{eqnarray}
&&\tilde j_{5\mu}^{(+)}(x,\epsilon)
\equiv\\\nonumber\\\nonumber
&&\bar\psi^+(x+\epsilon)\gamma_\mu\gamma_5\tau^3
e^{ieQ\int_x^{x+\epsilon}A(z)\cdot dz}\psi^+(x)
+\sigma^+(x)\partial_\mu\pi^{0+}(x)-\pi^{0+}(x)\partial_\mu\sigma^+(x)\;
. \\\nonumber
\end{eqnarray}
The Wilson-line has been introduced to render the point-split operator
gauge-invariant. This is the analogous to the treatment of the axial
anomaly in spinor QED for the {\em vacuum} \cite{hagen}.

The gauge-fields can be treated simply as background fields
without loss of generality. Using the equations of motion it is now easy to see
that 
\begin{eqnarray}
\langle\partial_\mu \tilde j^\mu_5(x,\epsilon)\rangle
=h\langle\pi^{0+}(x)\rangle+
ie\langle\bar\psi^{+}(x+\epsilon)\gamma^\mu\gamma_5
Q\tau_3\psi^+(x)\rangle F_{\mu\nu}\epsilon^\nu(1+O(\epsilon))
\label{awi1}\\\nonumber
\end{eqnarray}
The correlator on the right hand side of Eq. (\ref {awi1}) has short distance
singularities as 
$\epsilon^\mu
\rightarrow 0$, which we now
extract via a one-loop calculation. The question that we want to answer
is whether the nonequilibrium path integral introduces any new divergent
contributions to the Ward identity. At one-loop we find
\begin{eqnarray}
&&\langle\bar\psi^{+}(x+\epsilon)\gamma^\mu\gamma_5 Q\tau_3\psi^+(x)\rangle
=-ie \frac{N_c}{6}\int d^4z\left\{ 
\mbox{Tr}\left[S^{++}(z,x+\epsilon)\gamma^\mu\gamma_5S^{++}(x,z)
\gamma^\nu\right]+ 
\right.\label{axicorr}\\\nonumber\\\nonumber
&&\left.-\mbox{Tr}\left[S^{-+}(z,x+\epsilon)
\gamma^\mu\gamma_5S^{+-}(x,z)\gamma^\nu\right]\right\}A_\nu(z). 
\\\nonumber
\end{eqnarray}

Rewriting Eq. (\ref{axicorr}) in terms of the spatial Fourier transforms of
the Green's functions we obtain 

\begin{eqnarray}
-ie\frac{N_c}{6}&&\int dt^\prime\int \frac{d^3q}{(2\pi)^3}
e^{i\vec q\cdot x}\tilde A_{\nu}(\vec q, t^\prime)
\int\frac{d^3k}{(2\pi)^3}e^{-i\vec k \cdot \vec \epsilon}\times\label{axicorr1}
\\\nonumber\\\nonumber
&&\left\{\mbox{Tr}\left[S^{++}_k(t^\prime,t+\epsilon^0)
\gamma^\mu\gamma_5S^{++}_{k+q}(t,t^\prime)\gamma^\nu\right]
-\mbox{Tr}\left[S^{>}_k(t^\prime,t+\epsilon^0)
\gamma^\mu\gamma_5 S^{<}(t,t^\prime)_{k+q}\gamma^\nu\right]\right\}.
\\\nonumber
\end{eqnarray}   

There are {\em two} distinct contributions to the Ward identity
out-of-equilibrium.
The first one is obtained by contractions with fields on the forward
time contour and involves the usual time-ordered propagators. The second term
which arises from contractions with fields on the backward time contour is the
new contribution that appears in the nonequilibrium case.
Non-trivial modifications of the Ward identity can only arise from terms that
diverge as $\epsilon\rightarrow 0$. But the latter are short distance
divergences and therefore we need only to consider the UV behaviour of the
integrand in Eq. (\ref{axicorr1}). As $|\vec k|\rightarrow\infty$,
$g\sigma^2(t)$  
$g\dot\sigma(t)$ and the temperature $T_i$ can be ignored and the propagators in
Eqs. (\ref{sgredef}), (\ref{slessdef}) and (\ref{sgsl})   
take on their usual
zero temperature short distance forms, which is simply dictated by the singularities of free-field theory in the operator
product expansion of  fields. Hence the UV portion of the first term can be
rewritten in terms of the Fourier 
transformed fields as  
\begin{eqnarray}
\nonumber\\
ie\frac{N_c}{6}\int \frac{d^4q}{(2\pi)^4}e^{-iq.x}
\tilde A_\nu(q)\int \frac{d^4k}{(2\pi)^4}
e^{ik\cdot\epsilon}\frac{-4i\epsilon^{\alpha\mu\beta\nu} 
k_\alpha q_\beta}{k^2(k+q)^2}
\\\nonumber
\end{eqnarray}
which, after evaluating the divergent integral, yields
\begin{eqnarray}
\nonumber\\
\frac{-ieN_c}{12\pi^2}\left(\frac{\epsilon_\alpha}{\epsilon^2}
\right)
\partial_\beta A_\nu(x)\epsilon^{\alpha\mu\beta\nu}.
\\\nonumber
\end{eqnarray} 

So the contribution of this term to the Ward identity is then
\begin{eqnarray}
\nonumber\\
&&=\lim_{\epsilon\rightarrow 0}\left(\frac{\epsilon_\alpha\epsilon^\gamma}
{\epsilon^2}\right)\frac{e^2N_c}{24\pi^2}F_{\beta\nu}F_{\gamma\mu}
\epsilon^{\alpha\mu\beta\nu}\\
&&=-\frac{e^2N_c}{96\pi^2}F_{\mu\nu}F_{\alpha\beta}
\epsilon^{\mu\nu\alpha\beta}\label{forwardanom}
\\\nonumber
\end{eqnarray}

We now isolate the ultraviolet (the leading terms in the $\epsilon\rightarrow0$
limit) contributions to the second term in Eq. (\ref{axicorr1}) which
is due to the presence of the 
backward time contour 
\begin{eqnarray}
\nonumber\\
&&\frac{-ieN_c}{6}(4i\epsilon^{\alpha\mu\beta\nu})\times\\\nonumber
&&\times
\int\frac{d^4q}{(2\pi)^4}e^{i\vec q\cdot\vec x}
\tilde A_\nu(\vec q,\omega)\int \frac{d^3k}{(2\pi)^3}
\frac{e^{ik\cdot\epsilon}e^{it(|\vec k +\vec q|+k)}}
{4k|\vec k+\vec q|}
\int dt^\prime e^{it^\prime(\omega-|\vec k+\vec q|)-k)}k_\alpha p_\beta
\\\nonumber\\\nonumber
&&=\frac{ieN_c}{6}(4i\epsilon^{\alpha\mu\beta\nu})\\\nonumber
&&\times\int\frac{d^4q}{(2\pi)^4}e^{i\vec q\cdot\vec x}\tilde
A_\nu(\vec q,\omega) 
e^{i\omega t}\int \frac{d^3k}{(2\pi)^3}
\frac{e^{ik\cdot\epsilon}}{4k|\vec k +\vec q|}
2\pi \delta(\omega-|\vec k+\vec q|-k) k_\alpha p_\beta.
\\\nonumber
\end{eqnarray}
Clearly the $\vec k$- integral is cut off at large momenta by the delta
function 
and is finite in the limit $\epsilon \rightarrow 0$. Because of the
presence of $\epsilon^{\nu}$ multiplying the 
expression (see Eq. (\ref{awi1})), this term does not survive the $\epsilon
\rightarrow 0$ limit, hence  there will be no further anomalous contributions
to the Ward identity 
other than Eq. (\ref{forwardanom}). Therefore the anomalous Ward identity
is in fact {\bf unchanged} by equilibrium finite temperature or the
nonequilibrium state, as expected: 
\begin{eqnarray}
\nonumber\\
\langle\partial_\mu j^{+\mu}_5(x)\rangle=
h\langle \pi^{0+}\rangle-\frac{e^2N_c}{96\pi^2}
F_{\mu}F_{\nu}\epsilon^{\mu\nu\alpha\beta}.
\\\nonumber
\end{eqnarray}

That there are no corrections in or out of equilibrium should  come as
no surprise since the 
anomalous contributions to the axial Ward identity are generated by the 
ultraviolet (short distance) 
properties of correlators  and should not be affected by either temperature, finite density or the time dependence of
background fields or long-wavelength fluctuations. However, since a detailed
calculation in a medium in real time in or out of equilibrium was missing, this analysis provides a reassuring generalization of the vacuum result. 
The calculation of the expectation value for the backward branch follows
the same steps and leads to the same results. 

\section{Nonequilibrium Fermion Green Functions}
Below we outline the construction of the nonequilibrium Green's functions for
the fermions in the presence of the time-dependent background condensate
$\sigma_0(t)$ \cite{boyareheat}.
We  consider the solutions of the time dependent Dirac equation

\begin{equation}
[i\partial\!\!\!\slash-g\sigma_0(t)]\psi(\vec x,t)=0 
\end{equation}

Writing the four independent solutions as
\begin{eqnarray}
&&{\cal{U}}^{(1,2)}(\vec{x},t)= e^{i\vec{k}\cdot\vec{x}}U^{(1,2)}_k(t)\;,
\nonumber \\
&&{\cal{V}}^{(1,2)}(\vec{x},t)= e^{-i\vec{k}\cdot\vec{x}}V^{(1,2)}_k(t)\;,
\nonumber
\end{eqnarray}
the mode functions are found to satisfy 
\begin{eqnarray}
&&\left[i\gamma_0\frac{d}{dt}-\vec{\gamma}\cdot\vec{k}-g\sigma_0(t)\right]
U^{(1,2)}_k(t)=0\;, \label{uspinor} \\
&&\left[i\gamma_0\frac{d}{dt}+\vec{\gamma}\cdot\vec{k}-g\sigma_0(t)\right]
V^{(1,2)}_k(t)=0\;. \label{vspinor}
\end{eqnarray}
It turns out that it is convenient to write the spinors as
\begin{eqnarray}
&&U^{(1,2)}_k(t) =
\left[i\gamma_0\frac{d}{dt}-
\vec{\gamma}\cdot\vec{k}+g\sigma_0(t)\right]f_k(t)u^{(1,2)}\;,
\label{ffunct}\\
&&V^{(1,2)}_k(t) =
\left[i\gamma_0\frac{d}{dt}+
\vec{\gamma}\cdot\vec{k}+g\sigma_0(t)\right]g_k(t)v^{(1,2)}\;,
\label{gfunct}
\end{eqnarray}
with $u^{(1,2)}$, $v^{(1,2)}$ the spinor eigenstates of $\gamma_0$ with
eigenvalues $+1$, $-1$ respectively. The functions $f_k(t)$, $g_k(t)$ obey
the second order equations
\begin{eqnarray}
\left[\frac{d^2}{dt^2}+\vec{k}^2+g^2\sigma_0^2(t)-ig\dot{\sigma}_0(t)\right]
f_k(t)=0\;, \label{fequation} \\
\left[\frac{d^2}{dt^2}+\vec{k}^2+g^2\sigma_0^2(t)+ig\dot{\sigma}_0(t)\right]
g_k(t)=0\;. \label{gequation}
\end{eqnarray}
We now need to append initial conditions. We will consider the situation in
which the system was in equilibrium at a temperature $T_i$ for all times  
$t\leq t_i$ with the
expectation 
value of the scalar field being $\sigma(0)$ and $\dot{\sigma}(0)=0$.
 Thus the fermion mass is constant and
given by $g\sigma_0(0)$. We can now impose the condition that the modes
$f_k(t)$, $g_k(t)$ describe positive and negative frequency solutions for
$t\leq t_i$ and, normalizing the spinor solutions to the Dirac equation to unity, 
we impose the following initial conditions that describe particle and antiparticle states for $t<t_i<0$
\begin{eqnarray}
&&f_k(t<0)= \frac{e^{-i\omega_k t}}{\sqrt{2\omega_k(\omega_k
+g\sigma_0(0) ) }} \;,
\label{bcfoft}\\
&&g_k(t<0)= \frac{e^{i\omega_k t}}{\sqrt{2\omega_k(
2\omega_k+g\sigma_0(0) ) }} \;,
\label{bcgoft}\\
&& \omega_k= \sqrt{{\vec{k}}^2+g^2\sigma_0^2(0)}\;. \label{k0}
\end{eqnarray}
Equations (\ref{gequation}) with these boundary conditions imply that
\begin{equation}
g_k(t)=f^*_k(t)\;. 
\nonumber
\end{equation}
\subsection{Initial temperature $T_i=0$:}
The necessary ingredients for the zero temperature fermionic Green's
functions are the following
\begin{eqnarray}
&& S^>_{0k}(t,t')= -i \sum_{\alpha=1,2}U^{\alpha}_{\vec{k}}(t)
 \bar{U}^{\alpha}_{\vec{k}}(t')\;,
 \label{sgreat}  \\
&& S^<_{0k}(t,t')= i \sum_{\alpha=1,2}V^{\alpha}_{-\vec{k}}(t)
 \bar{V}^{\alpha}_{-\vec{k}}(t') \;. \label{sles}
\end{eqnarray}
In the standard Dirac representation for the $\gamma$ matrices, we find
\begin{eqnarray}
&&S^>_{0k}(t,t')= \\\nonumber
&&-i f_k(t)f^*_k(t')\left[{\cal{W}}_k(t)\gamma_0-
\vec{\gamma}\cdot
\vec{k}+g\sigma_0(t)\right]\left(\frac{1+\gamma_0}{2}\right)
\left[{\cal{W}}^*_k(t')\gamma_0-\vec{\gamma}\cdot
\vec{k}+g\sigma_0(t')\right]\;, \label{sgreatoft}\nonumber\\
&&S^<_{0k}(t,t')= \\\nonumber
&&-i f^*_k(t)f_k(t')\left[{\cal{W}}^*_k(t)\gamma_0-
\vec{\gamma}\cdot
\vec{k}-g\sigma_0(t)\right]\left(\frac{1-\gamma_0}{2}\right)
\left[{\cal{W}}_k(t')\gamma_0-\vec{\gamma}\cdot
\vec{k}-g\sigma_0(t')\right]\;,\nonumber\\
&&\quad {\cal{W}}_k(t)= i \frac{\dot{f}_k(t)}{f_k(t)}\;.
\label{timefreq}
\end{eqnarray}
With these basic results in place, the nonequilibrium fermionic Green's
functions can be 
constructed using Eqs. (\ref{ferplusmin}), (\ref{ferplusmin}),
(\ref{fertimeord}) and (\ref{ferantitimeord}). The fact that the equality
${\mbox{Tr\,}} S^{++}_{0k}(t,t)= {\mbox{Tr\,}} S^{--}_{0k}(t,t)=
{\mbox{Tr\,}} S^{>}_{0k}(t,t)= {\mbox{Tr\,}} S^{<}_{0k}(t,t)$ is
satisfied (where the trace is over Dirac indices) provides an important
check. The mode functions $f_k(t)$ 
satisfy yet another important property which is a consequence of probability
conservation and can also be checked explicitly\cite{boyareheat},
\begin{equation}
\left[-i\dot{f}^*_k(t)+g\sigma_0(t) f^*_k(t)\right]
\left[i\dot{f}_k(t)+g\sigma_0(t)f_k(t)\right]+k^2\,f^*_k(t)f_k(t)=1\;.
\end{equation}
\subsection{Initial temperature $T_i\neq0$:}

When the initial temperature is not zero, the CTP contour extends into the
complex plane with a leg that runs from $t_0$ to $t_0-i\beta$.. The fermionic
fields and Green functions then satisfy antiperiodic boundary conditions:
\begin{eqnarray}
S_k^>(t_0-i\beta,t)=-S_k^<(t_0,t)\label{kms}.
\end{eqnarray}
Incorporating these boundary conditions then leads to the following formulae
for the nonequilibrium fermion propagators at finite initial temperature:
\begin{eqnarray}
&&S_{\vec{k}}^{>}(t,t')
=S_{0\vec k}^>(t,t^\prime)\left[1-n_{f}(\omega_k)\right]-
S_{0\vec k}^<(t,t^\prime)n_{f}(\omega_k)\;,\label{sgredef1}
\\\nonumber\\
&&S_{\vec{k}}^{<}(t,t')= S_{0\vec k}^<(t,t^\prime)\left[1-n_f(\omega_k)\right]-
S_{0\vec k}^>(t,t^\prime)n_{f}(\omega_k)\;,
\label{slessdef1}\\\nonumber\\
&&\omega_k=\sqrt{\vec{k}^2+g^2\sigma^2(0)}\;,\quad\quad\quad
n_{f}(\omega_k)=\frac{1}{e^{\beta \omega_k}+1}\;. \label{fermifactor1}
\nonumber\\\nonumber
\end{eqnarray}
In equilibrium, i.e. when $\sigma_0$ is a constant function of time, the
fermion propagators are:
\begin{eqnarray}
&& S_{\vec{k}}^{>}(t,t')=
-\frac{i}{2\omega_k}\left[e^{-i\omega_k(t-t')}
(\not\!{k}+m_{\psi})
(1-n_{f}(\omega_k))+e^{i\omega_k(t-t')}\gamma_0
(\not{k}-m_{\psi})\gamma_0 n_{f}(\omega_k) \right]\;,
\nonumber\\
&& S_{\vec{k}}^{<}(t,t')=
\frac{i}{2\omega_k}\left[e^{-i\omega_k(t-t')}
(\not\!{k}+m_{\psi})
n_{f }(\omega_k)+e^{i\omega_k(t-t')}\gamma_0
(\not\!{k}-m_{\psi})\gamma_0 (1-n_{f}(\omega_k)) \right]\;.
\label{fermeqbm}\\\nonumber
\end{eqnarray}














\end{document}